\documentclass[sigconf,nonacm]{acmart}

\usepackage{tabularx}
\usepackage{longtable}
\usepackage{colortbl}
\setcopyright{none}
\hypersetup{pdflang={en},pdfauthor={Zihong He, Chen Liang, Hai-Ning Liang}}

\ExplSyntaxOn
\str_new:N \l_applooper_nav_contents_str
\NewDocumentCommand \AppLooperLinkContents {m}
  {
    \cs_if_exist:NT \pdfannot_dict_put:nne
      {
        \pdf_string_from_unicode:nnN {utf8/string} {#1}
          \l_applooper_nav_contents_str
        \pdfannot_dict_put:nne {link/GoTo} {Contents}
          {\l_applooper_nav_contents_str}
      }
  }
\ExplSyntaxOff
\newcommand{\AppLooperFigureLink}[6]{%
  \makebox[0pt][l]{\hspace*{#1\unitlength}%
    \raisebox{#2\unitlength}[0pt][0pt]{%
      \begingroup
        \AppLooperLinkContents{#6}%
        #5{\makebox[#3\unitlength][l]{\rule{0pt}{#4\unitlength}}}%
      \endgroup}}%
  \ignorespaces
}

\newenvironment{AppLooperPrompt}{%
  \list{}{\leftmargin=1em\rightmargin=0pt\topsep=3pt\partopsep=0pt\parsep=0pt\listparindent=0pt}%
  \item\relax\small\itshape\emergencystretch=2em
  \let\AppLooperUnderscore\_%
  \renewcommand{\_}{\AppLooperUnderscore\allowbreak}%
  \textup{\textbf{Prompt:}}\space\ignorespaces}{\endlist}

\begin{document}

\title[AppLooper]{AppLooper: A Multi-Role Agentic Application Development Loop for Owner Participation and Release Trust}

\author{Zihong He}
\email{zhe154@connect.hkust-gz.edu.cn}
\affiliation{%
  \institution{HKUST (GZ)}
  \city{Guangzhou}
  \country{China}
}
\author{Chen Liang}
\email{chenliang2@hkust-gz.edu.cn}
\affiliation{%
  \institution{HKUST (GZ)}
  \city{Guangzhou}
  \country{China}
}
\author{Hai-Ning Liang}
\email{hainingliang@hkust-gz.edu.cn}
\affiliation{%
  \institution{HKUST (GZ)}
  \city{Guangzhou}
  \country{China}
}
\affiliation{%
  \institution{The Hong Kong University of Science and Technology}
  \city{Hong Kong}
  \country{China}
}
\renewcommand{\shortauthors}{He et al.}

\begin{abstract}
Coding agents based on large language models can autonomously implement, test, and revise applications across successive rounds. Yet checking of each version is often limited to the development agent's own reports and the application owner's manual inspection, which constrains the problems found before release and leaves owners with limited grounds for trusting the application they release. AppLooper is a development loop in which virtual-user, testing, and owner-intent simulation agents check each candidate version of the application for target-user experience, technical defects, and consistency with the owner's stated requirements, alongside the owner's hands-on trials. Their findings and the owner's feedback drive revision and retesting, and each finding, revision, and retest is recorded with the candidate version it concerns, giving the owner inspectable evidence for trusting the application before making the release decision. In a paired study with 30 participants, overall application preference and self-reported trust at the release decision favored AppLooper over an iterative Claude Code baseline with independent evaluation (20 versus 3 and 16 versus 3, respectively, with Holm-adjusted \(p\) values of .0020 and .0133). Open-ended responses showed that participants attended to whether additional features met their needs and whether development progress and version changes remained visible while waiting.
\end{abstract}




\begin{teaserfigure}
  \centering
  \includegraphics[width=.9\textwidth,alt={An application owner moves from an initial idea through a loop of development, virtual-user trials, testing, and owner-intent checks to human-authorized release. Each role has a distinct responsibility.}]{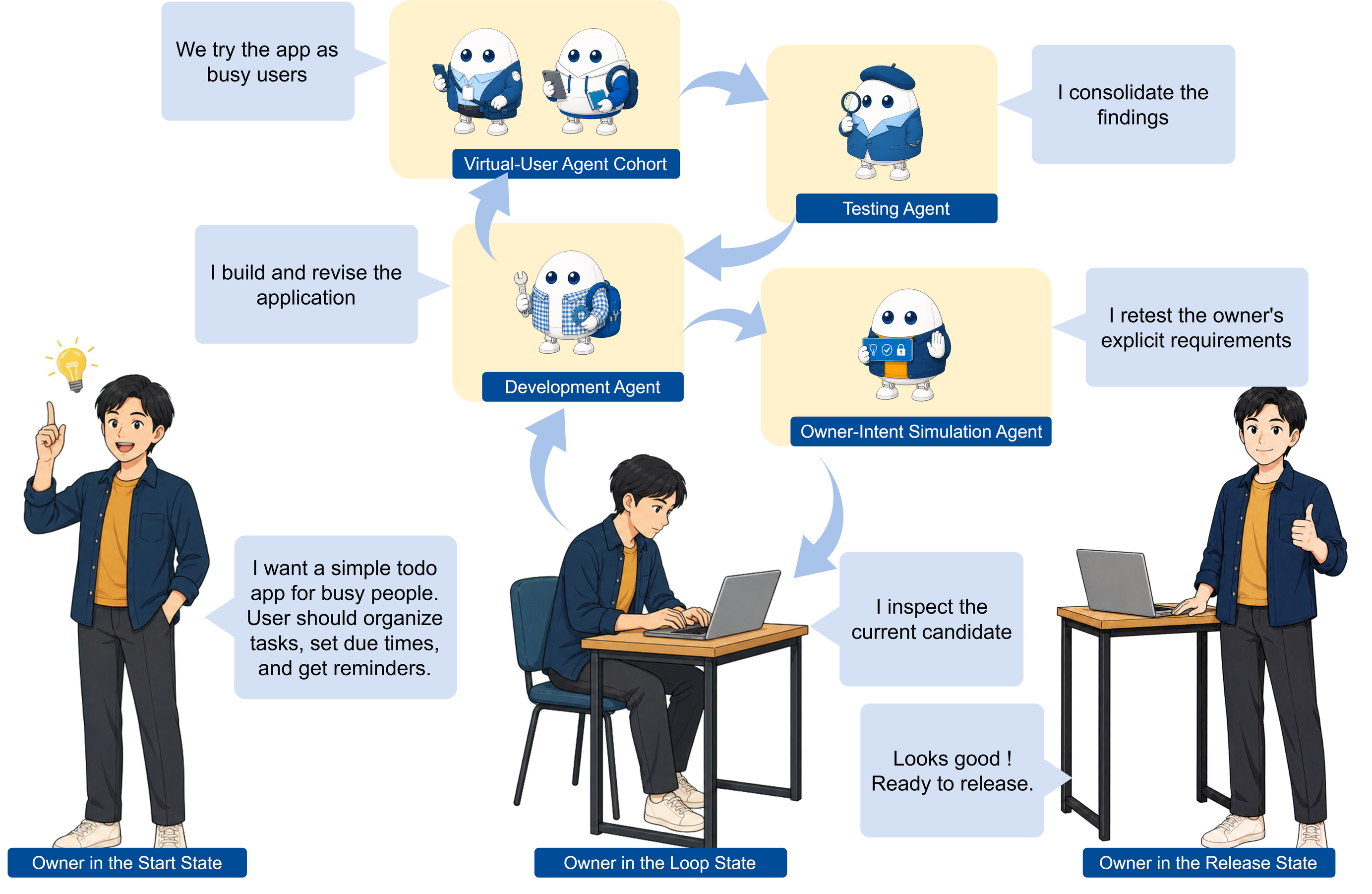}
  \caption{From initial intent to release, owners try candidate versions, submit feedback, and inspect revisions. The Development Agent revises candidates using evidence from the Virtual-User Agent Cohort, Testing Agent, and Owner-Intent Simulation Agent.}
  \label{fig:applooper-leading}
  \Description{The application owner starts with an idea, inspects candidate applications in an isolated trial window, can send issues to the development agent, and ultimately authorizes release. Within the loop, the development agent revises the application, virtual-user agents try it, the testing agent performs read-only regression and browser checks, and the owner-intent simulation agent prechecks feedback against the owner's own statements before development and retests relevant scenarios after revisions.}
\end{teaserfigure}

\maketitle

\section{Introduction}

Large language models (LLMs) increasingly support coding agents that use development tools to autonomously implement, test, and revise software in iterative cycles~\cite{liu2026large}. \citet{jimenez2024swe} show that resolving real repository issues often requires coordinated changes across functions, classes, and files. To support such work, \citet{yang2024swe} develop interfaces for coding agents to navigate repositories, edit files, and execute tests, while \citet{wang2025openhands} combine these capabilities with browser interaction and agent coordination. \citet{liu2026dive} describe a coding-agent execution loop of repeated model calls and tool execution, with returned results guiding subsequent actions.

Autonomous development also raises concerns about whether successive changes continue to satisfy users' requirements. \citet{takerngsaksiri2025human} report that engineers using an agent-based workflow with human review of coding plans and generated code perceived reduced effort for simple tasks while retaining concerns about code quality. \citet{tang2026coding} find that coding agents can report completion despite missing functionality or unmet requirements, and 91.49\% of visibly resolved misalignment episodes were corrected by agents after explicit developer pushback. \citet{wang2024investigating} further document how limited support for evaluating generated suggestions leaves developers relying on time-consuming manual checks to judge trustworthiness.

Verification becomes harder to sustain when users leave an agent running and return later. In a study of computer-use agents, \citet{grunde2026overseeing} find that lengthy, repetitive execution traces make returning to ongoing work cumbersome. Usability simulations offer another source of feedback about how an application supports user tasks. \citet{lu2025uxagent} simulate website interactions and report that user experience researchers value the resulting early feedback while questioning how well the simulated behavior represents real users. Coding agents can continue revision across successive rounds~\citep{liu2026large,liu2026dive}. When an application owner, the person who commissions an application and decides on its release, returns after leaving the agent running, the owner must work out from a lengthy, repetitive record what has changed in the meantime~\citep{grunde2026overseeing} and needs additional time to assess the current application and determine the next action~\citep{endsley1995out}. An agent may report task completion when required functionality is still missing~\citep{tang2026coding}. Support for evaluating that report remains limited, and trust in the result continues to depend on time-consuming manual inspection~\citep{wang2024investigating}. These findings indicate that, as coding agents continue to revise and refine an application across rounds, an owner's trust in the application to be released depends on manual inspection, and owners have limited support for the three steps this inspection involves, namely working out what has changed in their absence, assessing the current application, and judging whether a reported completion is accurate.

Motivated by these limitations, we introduce \textbf{AppLooper}, a multi-role agentic application development loop for owner participation and feedback from hands-on use (Figure~\ref{fig:applooper-leading}). A candidate version is the application's code as it stands at a given moment, and every change to the code yields a new one. Owners run and try each candidate version and comment on their experience. Development, virtual-user, testing, and owner-intent simulation agents contribute checks and revisions that owners can inspect before release and on which they can base their trust in the application.

We introduce multiple agents associated with the application development process to check and revise each candidate from several perspectives and to give application owners evidence, recorded with the version it was collected on, for working out what has changed, assessing the current application, and judging reported completion during continued revision. Their work covers implementation and revision, target-user experience checks, technical verification, and checks confined to owner-stated requirements. The development agent uses Claude Code to implement the initial intent and subsequent owner feedback as a revised application. Virtual-user agents execute browser tasks under target-user personas and report experience findings before owner inspection. The testing agent performs read-only failure reproduction, regression testing, and interface inspection, and does not modify the product. The owner-intent simulation agent examines incoming feedback against only the requirements, constraints, and feedback stated or confirmed by the owner, and retests the corresponding scenarios after revision. Each finding keeps the role that produced it and its candidate version, so later inspection can trace a result to the version it concerns. A finding that a saved task shows no deadline, for example, is stored as a virtual user's report on the candidate version in which it appeared.

AppLooper follows a \textbf{feedback, revision, retesting, and acceptance} process and records each step with the candidate version it concerns, so an owner can follow a comment to the revision it prompted and to the later check of the revised version. Owner feedback is preserved verbatim and stored separately from agent-generated assessments. Owners inspect executable candidates in isolated trial windows, report issues, and review revisions through targeted inspection scenarios. Each scenario presents the reported problem, relevant task and interface evidence, version changes, and the owner's next check. In an example to-do application, the owner requested a Today filter and a selectable reminder interval, and specified that the request was satisfied only if a task due today remained under Today after reload and its card showed the chosen interval. The development agent added the filter and an interval selector, and the owner-intent simulation agent retested the revised candidate against these conditions, finding the Today filter present and the existing filters working (Figure~\ref{fig:applooper-conversation-example}). The owner could then check the same conditions on that candidate. Progress displays show the current stage, requirement status, and next actions. The owner reviews completed checks and unresolved issues before making the final release decision. Final approval applies only to the candidate version the owner inspected, which stays unchanged during inspection, and any later change requires renewed acceptance.

To examine the incremental benefit of AppLooper's multi-role structure relative to an iterative coding workflow that already separates implementation from independent evaluation, we compared the two in a paired study with 30 participants. Our baseline, Claude Code Loop, follows the evaluator and optimizer pattern~\citep{anthropic2024effectiveagents}. A development agent uses the Claude Code model and tool loop to implement the application and to revise it from the owner's feedback and the evaluator's findings~\citep{liu2026dive}. A separate evaluator checks the candidate against the frozen requirement (the requirement statement the owner confirmed before development), critical user journeys, and layered tests, and returns those findings for revision. AppLooper retains this development agent, this evaluator, the same coding model and initial requirements, the same trial and feedback interface, a 20-round cap, and the same owner acceptance procedure. AppLooper adds three roles that check each candidate version. Virtual-user agents carry out the target-user tasks and report experience findings before owner inspection. The testing agent reproduces failures and runs regression without modifying the product. The owner-intent simulation agent checks new feedback against only the requirements stated by the owner and retests the corresponding scenarios after revision. Orchestration, the component that coordinates the agents, records each finding, the revision it prompted, and the later retest with the candidate version they concern.

In this comparison, 20 participants preferred their AppLooper application, 3 preferred the Claude Code Loop application, and 7 rated them similarly. At the release decision, 16 participants reported greater trust in the AppLooper application and 3 in the baseline. Both comparisons favored AppLooper after Holm correction~\citep{holm1979simple} ($p_H=.0020$ and $.0133$). Mean ratings of final application quality (3.80 versus 3.17), effort saved when using the finished application (3.80 versus 2.87), and traceability of the released version (3.67 versus 3.17) were also higher for AppLooper in unadjusted paired tests ($p=.0072$, $.0051$, and $.0169$). Owner-intent prechecks or retests accompanied owner feedback in 23 of the 30 AppLooper workflows. Some participants valued the effort-saving checks from several roles, and some valued seeing how and why each change was made and which feedback it addressed.

This work makes three contributions.

\begin{enumerate}
    \item We propose and implement AppLooper, a multi-role agentic application development system that supports owners in confirming the frozen requirement before iteration, trying candidate versions and reporting issues at development checkpoints, and approving, returning, or deferring release when autonomous iteration pauses. Its core code implementation is available at \url{https://github.com/ZihongHe/applooper}.
    \item A mechanism that records each piece of feedback, revision, retest, and acceptance decision with the candidate version it concerns. Owner feedback is recorded separately from simulated evidence, the results produced by agents that simulate users or the owner, so the agents can continue revision across rounds while owners can inspect each candidate's development artifacts and assess its trustworthiness before release.
    \item A paired study with 30 participants. After Holm correction~\citep{holm1979simple}, 20 of the 30 preferred the AppLooper application ($p_H=.0020$) and 16 reported greater trust in it at the release decision ($p_H=.0133$). These results indicate that owners preferred the applications produced with AppLooper and trusted them more when authorizing release.
\end{enumerate}

\section{Related Work}

This section reviews work on coding agents and feedback-driven loops, human participation and verification, and virtual users.

\subsection{LLM-Based Coding Agents and Feedback-Driven Loops}

Code completion leaves developers responsible for operating the editor and assessing local suggestions. \citet{mozannar2024reading} identify interaction and time costs from prompting, inspecting, editing, and validating these suggestions. Coding agents move interaction toward task specification and artifact inspection. \citet{yang2024swe} frame agents as computer users requiring dedicated interfaces, and \citet{wang2025openhands} combine sandboxes, command lines, browsers, and agent coordination in a general development platform. \citet{barke2023grounded} distinguish acceleration, when programmers know their next step, from exploration of possible directions. Users may prefer generation tools despite limited efficiency gains and difficulty understanding or debugging generated code \citep{vaithilingam2022expectation}.

\citet{yao2022react} interleave reasoning, actions, and observations so new evidence informs subsequent actions. Code execution, test results, and environmental state provide observable grounds for adjustment. Iterative methods reuse information from failed attempts. \citet{madaan2023self} repeatedly generate feedback and revise outputs with the same model, while \citet{shinn2023reflexion} store verbal reflections in episodic memory to inform later attempts. For programming, \citet{chen2024teaching} combine execution with natural-language debugging explanations, and \citet{huang2023agentcoder} use specialized agents for iterative testing and optimization.

Role specialization also structures software production. \citet{qian2024chatdev} coordinate design, coding, and testing, and \citet{hong2024metagpt} encode development procedures as role-specific prompts. Superpowers\footnote{\url{https://github.com/obra/superpowers} (accessed 15 November 2026).} organizes requirements clarification, planning, test-driven development, and code review. AppLooper organizes its loop around owner trials and records, for each piece of feedback from use, the candidate version it concerns and the revision and retest that follow.

\subsection{Human Participation and Verification in Agentic Development}

User-centered design places users throughout development. \citet{gould1985designing} advocate early, continuous attention to users, empirical measurement through actual use, and iterative design informed by evaluation. \citet{kujala2003user} finds that participation supports requirement elicitation and satisfaction, although participant roles, tacit needs, and costs require consideration. \textit{User participation} denotes activities during development, and \textit{user involvement} denotes the system's perceived importance and personal relevance \cite{barki1989rethinking,barki1994measuring}. Message counts, clicks, or final satisfaction alone therefore cannot establish meaningful engagement.

\citet{takerngsaksiri2025human} introduce Human-in-the-loop LLM-based Agents (HULA), letting engineers refine plans, review generated code, and guide agents across stages. Engineers reported reduced planning and coding effort for some simple tasks but retained quality concerns. This participation primarily involves professional developers inspecting plans and source code. \citet{yan2024ivie} support inspection through lightweight explanations beside generated code, while \citet{mozannar2024reading} document the interaction and time costs of reading, judging, and validating AI-generated code.

In a navigation experiment, \citet{endsley1995out} found that operators who had stayed less involved in the ongoing control understood the situation less well and took longer to decide after a failure. Greater retained operator control shortened that delay. Studying computer-use agent supervision, \citet{grunde2026overseeing} found that returning participants relied on summaries and process records but encountered lengthy, repetitive records. An interface linking requirements, agent assumptions, and execution records shortened error-finding time and increased confidence without clearly improving overall judgment accuracy. These findings motivate concise support for returning users to regain an overview of the work, while cautioning against equating confidence with verification accuracy.

Through interviews with 17 developers, \citet{wang2024investigating} identify appropriate expectations, AI-tool configuration, and suggestion verification as central trust challenges. Human and AI interaction guidelines likewise emphasize communicating capabilities, supporting correction, and enabling understanding and control \citep{amershi2019guidelines}. These principles motivate showing returning users the state of the current version, what the available evidence covers, known limitations, and the actions open to the owner. \citet{dhanorkar2026human} identify a priori control, co-planning, real-time monitoring, and post hoc review as complementary forms of oversight. Yet agents can misinterpret user goals and rules. \citet{tang2026coding} operationalize misalignment through explicit developer objections and find that 91.49\% of visibly resolved cases required explicit correction. Their observations span project understanding, intent interpretation, rule compliance, action scope, implementation and execution, and progress reporting. Throughout collaboration, it remains necessary to keep track of how the current version relates to the owner's goals and feedback.

Requirements traceability is the ability to follow a requirement back to its origins and stakeholder context and forward to the development artifacts that implement it \cite{gotel1994analysis}. Broader research emphasizes creating, maintaining, and using trusted links among evolving artifacts, noting that ad hoc retrospective tracing often fails to deliver intended benefits \cite{cleland2014software}. AppLooper applies these principles across development and release. Owners identify target users and core problems, confirm the frozen requirement, inspect executable candidates, and submit feedback. Each agent check and revision is recorded with the performing role and the version it applies to. AppLooper's orchestration and state management records which tasks and subsequent checks follow from each group of related feedback. Suppose a virtual user reports that a saved task shows no deadline. The record then shows the development agent's revision and the same virtual user's retest, each with the version on which it was made. Intermediate acceptance through actual interface use, accepted feedback, and unresolved risks inform final release decisions. In this loop every check and approval refers to one candidate version, so the owner authorizes release on the basis of checks made on that same version.

\subsection{Virtual Users and Formative User Experience Evaluation}

Real-user testing may not keep pace with rapid prototyping, motivating simulated interactions as early formative signals, feedback collected during development to guide revision. \citet{xiang2024simuser} simulate mobile application use from user characteristics, contexts, and tasks and generate heuristic usability feedback. For a simple smartwatch interface, similarity between issues identified by SimUser and real users ranged from 35.7\% to 100\% across groups and usability categories. \citet{lu2025uxagent} combine a persona generator, LLM agent module, universal browser connector, and results interface to generate and inspect simulated website interactions. In their evaluation with 16 user experience researchers, participants recognized exploratory value while expressing concerns about future LLM use in user experience research.

Long-horizon browser tasks remain challenging. WebArena reports a substantial task-success gap between its strongest GPT-4 baseline and humans on reproducible, functional websites \citep{zhou2024webarena}. Successful execution thus offers task-specific evidence whose relevance to human experience requires real-user evaluation.

Representing variation within target populations also matters. \citet{bui2025mixture} combine weighted personas and examples to improve synthetic-data distributional alignment and diversity. \citet{liu2026beyond} model populations through prevalence-weighted behavioral groups and show that individually plausible conversations can distort aggregate behavior. These findings support documenting cohort composition, weights, and the data behind the personas, and claiming representativeness only for a justified reference distribution.

\citet{seshadri2026lost} find that user-LLM choice affects estimated agent performance, calibration varies with task difficulty, and proxy validity can vary across languages and populations. Although their study concerns conversational benchmark users rather than application usability, it underscores the need to interpret simulation relative to models, populations, and tasks. Together with varying SimUser coverage and researchers' concerns, this positions virtual users as a source of formative evidence that applies within the stated models, populations, and tasks.

AppLooper applies such simulated use~\citep{xiang2024simuser,lu2025uxagent} to each candidate version during continued development. In line with work on persona populations~\citep{bui2025mixture,liu2026beyond}, each persona states the target-user population it covers, and each finding records which persona reported it. For example, in a to-do application, a finding that a deadline cannot be chosen on a phone could be reported by the virtual user whose persona is an office worker using a phone browser. Consistent with the studies above, these findings serve as formative evidence. They are labeled as simulated and recorded with the candidate version and the pages or controls they concern, and the owner tries the relevant scenarios in the application before deciding on release.

\section{AppLooper Roles}
\label{sec:roles}

\begin{figure*}[!t]
  \centering
  \begingroup
  \hypersetup{hidelinks}
  \setlength{\unitlength}{.01\linewidth}
  \makebox[\linewidth][l]{%
    \includegraphics[width=\linewidth,alt={AppLooper connects the Application Owner with feasibility and initialization and with candidate acceptance as the bounding stages, with the guardrail management and orchestration that hold during development between them, and with four agent roles. Solid arrows show task and artifact flows, while dashed arrows show feedback and evidence flows. Visible section references identify the containing section and link to the corresponding description. Internal agent modules link to the matching bolded terms within the role subsections. Only the Development Agent modifies product source. The owner retains release authority.}]{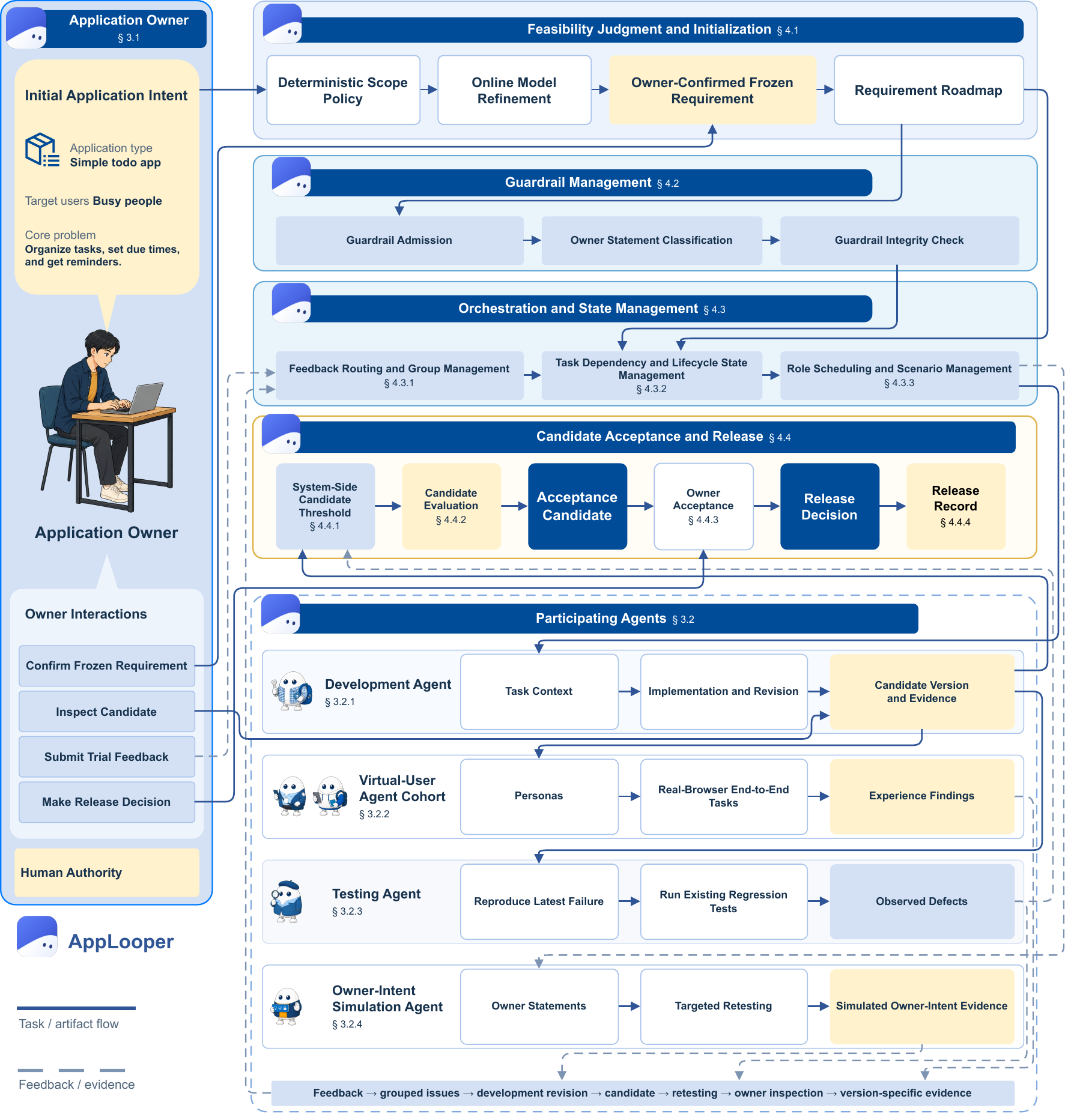}%
    \kern-\linewidth%
    \AppLooperFigureLink{1.195804}{99.839156}{18.041958}{3.524476}{\hyperref[sec:owner-participation]}{Application Owner: Section 3.1}
    \AppLooperFigureLink{24.608392}{100.342653}{70.741259}{2.349650}{\hyperref[sec:app_accepted]}{Feasibility Judgment and Initialization: Section 4.1}
    \AppLooperFigureLink{25.447552}{86.018177}{55.804196}{2.517483}{\hyperref[sec:guardrails]}{Guardrail Management: Section 4.2}
    \AppLooperFigureLink{25.447552}{74.286709}{55.804196}{2.517483}{\hyperref[sec:lifecycle-orchestration]}{Orchestration and State Management: Section 4.3}
    \AppLooperFigureLink{25.447552}{62.110485}{69.146853}{2.937063}{\hyperref[sec:candidate-acceptance]}{Candidate Acceptance and Release: Section 4.4}
    \AppLooperFigureLink{25.447552}{45.285310}{60.083916}{2.769231}{\hyperref[sec:agent-roles]}{Participating Agents: Section 3.2}
    \AppLooperFigureLink{1.783217}{44.303492}{16.363636}{5.664336}{\hyperref[sec:owner-participation]}{Owner Interactions: Section 3.1}
    \AppLooperFigureLink{1.363636}{20.739856}{17.202797}{4.531469}{\hyperlink{module:human-authority}}{Human Authority: Section 3.1, module description}
    \AppLooperFigureLink{24.860140}{92.714681}{14.265734}{6.419580}{\hyperlink{module:deterministic-scope-policy}}{Deterministic Scope Policy: Section 4.1, module description}
    \AppLooperFigureLink{40.804196}{92.714681}{14.265734}{6.419580}{\hyperlink{module:online-model-refinement}}{Online Model Refinement: Section 4.1, module description}
    \AppLooperFigureLink{56.748252}{92.714681}{19.300699}{6.419580}{\hyperlink{module:owner-confirmed-frozen-requirement}}{Owner-Confirmed Frozen Requirement: Section 4.1, module description}
    \AppLooperFigureLink{77.727273}{92.714681}{17.622378}{6.419580}{\hyperlink{module:requirement-roadmap}}{Requirement Roadmap: Section 4.1, module description}
    \AppLooperFigureLink{25.699301}{79.615380}{23.076923}{4.531469}{\hyperlink{module:guardrail-admission}}{Guardrail Admission: Section 4.2, module description}
    \AppLooperFigureLink{50.454545}{79.615380}{20.559441}{4.531469}{\hyperlink{module:owner-statement-classification}}{Owner Statement Classification: Section 4.2, module description}
    \AppLooperFigureLink{72.692308}{79.615380}{22.237762}{4.531469}{\hyperlink{module:guardrail-integrity-check}}{Guardrail Integrity Check: Section 4.2, module description}
    \AppLooperFigureLink{25.699301}{67.044751}{23.076923}{4.531469}{\hyperref[sec:feedback-clusters]}{Feedback Routing and Group Management: Section 4.3.1}
    \AppLooperFigureLink{50.454545}{67.044751}{20.559441}{4.531469}{\hyperref[sec:task-lifecycle-state]}{Task Dependency and Lifecycle State Management: Section 4.3.2}
    \AppLooperFigureLink{72.692308}{67.044751}{22.237762}{4.531469}{\hyperref[sec:role-scheduling]}{Role Scheduling and Scenario Management: Section 4.3.3}
    \AppLooperFigureLink{25.699301}{53.089506}{9.230769}{8.055944}{\hyperref[sec:system-candidate-threshold]}{System-Side Candidate Threshold: Section 4.4.1}
    \AppLooperFigureLink{37.447552}{53.089506}{9.230769}{8.055944}{\hyperref[sec:candidate-evaluation]}{Candidate Evaluation: Section 4.4.2}
    \AppLooperFigureLink{49.195804}{53.089506}{9.230769}{8.055944}{\hyperlink{module:immutable-acceptance-version}}{Acceptance Candidate: Section 4.4.2, module description}
    \AppLooperFigureLink{60.944056}{53.089506}{9.230769}{8.055944}{\hyperref[sec:owner-acceptance]}{Owner Acceptance: Section 4.4.3}
    \AppLooperFigureLink{72.692308}{53.089506}{9.230769}{8.055944}{\hyperlink{module:human-authorized-final-release}}{Release Decision: Section 4.4.4, module description}
    \AppLooperFigureLink{84.440559}{53.089506}{9.230769}{8.055944}{\hyperref[sec:accountable-release-record]}{Release Record: Section 4.4.4}
    \AppLooperFigureLink{29.727273}{36.725870}{13.426573}{6.251748}{\hyperref[sec:development-agent]}{Development Agent: Section 3.2.1}
    \AppLooperFigureLink{42.902098}{36.348247}{14.685315}{7.006993}{\hyperlink{module:bounded-development-context}}{Task Context: Section 3.2.1, module description}
    \AppLooperFigureLink{59.685315}{36.348247}{15.524476}{7.006993}{\hyperlink{module:implementation-revision}}{Implementation and Revision: Section 3.2.1, module description}
    \AppLooperFigureLink{77.307692}{36.348247}{17.202797}{7.006993}{\hyperlink{module:versioned-development-evidence}}{Candidate Version and Evidence: Section 3.2.1, module description}
    \AppLooperFigureLink{32.916084}{26.949646}{9.986014}{6.251748}{\hyperref[sec:virtual-user-cohort]}{Virtual-User Agent Cohort: Section 3.2.2}
    \AppLooperFigureLink{42.902098}{26.572023}{14.685315}{7.006993}{\hyperlink{module:explicit-personas}}{Personas: Section 3.2.2, module description}
    \AppLooperFigureLink{59.685315}{26.572023}{15.524476}{7.006993}{\hyperlink{module:browser-end-to-end-tasks}}{Real-Browser End-to-End Tasks: Section 3.2.2, module description}
    \AppLooperFigureLink{77.307692}{26.572023}{17.202797}{7.006993}{\hyperlink{module:candidate-bound-findings}}{Experience Findings: Section 3.2.2, module description}
    \AppLooperFigureLink{30.398601}{17.173422}{13.594406}{6.251748}{\hyperref[sec:testing-agent]}{Testing Agent: Section 3.2.3}
    \AppLooperFigureLink{42.902098}{16.795800}{14.685315}{7.006993}{\hyperlink{module:reproduce-latest-failure}}{Reproduce Latest Failure: Section 3.2.3, module description}
    \AppLooperFigureLink{59.685315}{16.795800}{15.524476}{7.006993}{\hyperlink{module:existing-regression-tests}}{Run Existing Regression Tests: Section 3.2.3, module description}
    \AppLooperFigureLink{77.307692}{16.795800}{17.202797}{7.006993}{\hyperlink{module:observed-defects}}{Observed Defects: Section 3.2.3, module description}
    \AppLooperFigureLink{30.398601}{7.397198}{12.209790}{6.251748}{\hyperref[sec:owner-intent-agent]}{Owner-Intent Simulation Agent: Section 3.2.4}
    \AppLooperFigureLink{42.902098}{7.019576}{14.685315}{7.006993}{\hyperlink{module:bounded-owner-evidence}}{Owner Statements: Section 3.2.4, module description}
    \AppLooperFigureLink{59.685315}{7.019576}{15.524476}{7.006993}{\hyperlink{module:targeted-retesting}}{Targeted Retesting: Section 3.2.4, module description}
    \AppLooperFigureLink{77.307692}{7.019576}{17.202797}{7.006993}{\hyperlink{module:simulated-owner-evidence}}{Simulated Owner-Intent Evidence: Section 3.2.4, module description}
  }%
  \endgroup
  \caption{AppLooper's application development loop. Feasibility and initialization and candidate acceptance bound the loop; the guardrail management and orchestration hold during development between them. Solid arrows show task and artifact flows. Dashed arrows show feedback and evidence flows. Section links lead to module descriptions.}
  \label{fig:framework}
  \Description{AppLooper connects the Application Owner with feasibility and initialization and with candidate acceptance as the bounding stages, with the guardrail management and orchestration that hold during development between them, and with four agent roles. Solid arrows show task and artifact flows, while dashed arrows show feedback and evidence flows. Visible section references identify the containing section and link to the corresponding description. Internal agent modules link to the matching bolded terms within the role subsections. Only the Development Agent modifies product source. The owner retains release authority.}
\end{figure*}

AppLooper lets owners confirm the requirement, try each candidate version, report issues, and decide on release~\citep{gould1985designing,kujala2003user}. It records owner feedback and agent checks, revisions, and retests on the candidate version they concern, so that owners can trace a requirement or feedback item to its revision and evidence~\citep{gotel1994analysis} and the release decision applies to the version they inspected (Figure~\ref{fig:framework}). The workflow first fixes the owner's intent as a frozen requirement and plans it as a Requirement Roadmap of confirmed constraints and tasks, then repeats agent checks, revision, and retesting on successive candidate versions until one is presented for the owner's acceptance. This section introduces the roles that take part, the application owner and four agent roles, and Section~\ref{sec:workflow} presents the workflow. ChatGPT (OpenAI) assisted with building the system code.

Sections~\ref{sec:agent-roles} and~\ref{sec:workflow} present the English versions of the role templates and principal call instructions beside the roles and steps they drive. For agent calls, the role template is the system prompt, and the call instruction is sent with its inputs as JSON, a structured data format. Most prompts are abridged, omitting operational clauses on preview, process, and workspace handling, test accounts and login forms, contact with users, external publication, interface-message wording, and output format. In the prompts, a guardrail is a confirmed constraint (Section~\ref{sec:guardrails}), Experience Twin is the runnable browser preview, Computer Use is operation of the real browser interface, CUJ is a critical user journey, and DAG is a directed acyclic graph of task dependencies.

\subsection{Application Owner}
\label{sec:owner-participation}

In the AppLooper system, the application owner takes part throughout development and holds the authority to release the application. The owner submits an initial intent comprising target users, application type, and core problem, and then acts through four interactions, confirming the frozen requirement, inspecting candidate versions, reporting trial issues, and making release decisions. Figure~\ref{fig:framework} shows where each interaction enters the system. For a to-do application, the owner confirms reminders before deadlines as part of the frozen baseline, opens each candidate through the runnable trial entry to try creating a task, reports that a task due today is missing under the Today filter, and later approves, returns, or defers that candidate in Owner Acceptance. State management records each operation's interface target (the page or control involved), interaction outcome, and candidate version.

At development checkpoints, an isolated trial window lets the owner exercise the current candidate directly and send observed issues to the development agent through its feedback entry point, the input box at the lower left of Figure~\ref{fig:applooper-pc-mobile}. The system preserves the original text, records the candidate version and interface target it refers to, and triggers a precheck against the owner's own statements, followed by revision and retesting. The record of that report holds the sentence as written, the Today filter as the interface target, and the candidate version being tried. 

\hypertarget{module:human-authority}{}Once the candidate meets the system-side threshold, the system's own readiness evaluation (Section~\ref{sec:system-candidate-threshold}), autonomous iteration pauses for a human decision. The owner performs core tasks in the isolated trial window, following a checklist generated for that candidate, and chooses to approve, return, or defer. The owner's own actions, inspection results, and decisions are stored separately from simulated evidence, which is labeled as such.

\subsection{Participating Agents}
\label{sec:agent-roles}

\begin{figure*}[!t]
  \centering
  \includegraphics[width=\linewidth,alt={Six chronological speech-bubble messages, with virtual-user feedback, testing, and a scope-confirmation request on the left, followed by owner confirmation, development changes, and retesting on the right.}]{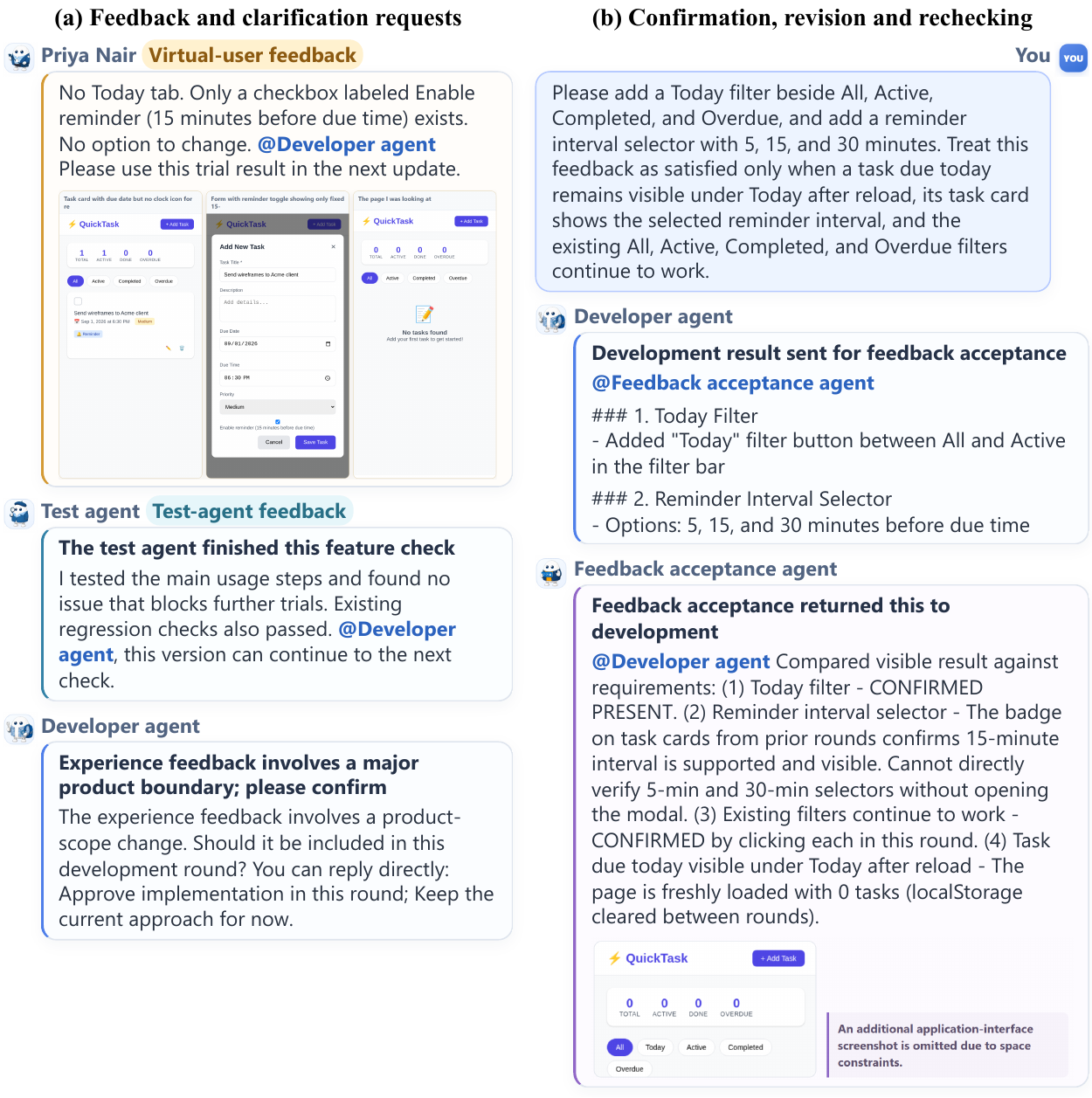}
  \caption{An excerpt from the development dialogue among agents and the owner of an example to-do application. (a) Feedback and requirement clarification. (b) Owner requirement confirmation, revision, and rechecking. Messages are ordered chronologically, reading the left panel before the right and each panel from top to bottom. The interface labels Developer agent, Test agent, and Feedback acceptance agent correspond to the Development Agent, Testing Agent, and Owner-Intent Simulation Agent, respectively. Redundant text and nonessential implementation details are omitted to keep the key interactions legible.}
  \label{fig:applooper-conversation-example}
  \Description{A development dialogue for a to-do application in six speech bubbles, read down the left panel and then down the right. The left panel contains virtual-user feedback with three application screenshots, passing testing feedback, and a development-agent request for scope confirmation. The right panel contains the owner's complete request for Today filtering and selectable reminder intervals, a development report, and feedback returned to development after retesting. The retest confirms the Today control and existing filters, reports that 5- and 30-minute choices cannot be directly verified, and describes a freshly loaded empty task list after localStorage was cleared between rounds. The retest's original application screenshot and screenshot-omission note remain visible.}
\end{figure*}

Four agent roles take part in development and separate revision, target-user experience evaluation, technical verification, and evaluation against owner-stated requirements. Distinct role avatars, shown beside each agent's name in Figure~\ref{fig:applooper-conversation-example}, identify agents in the interface. Each role sees only the requirements, tasks, candidate version, and feedback its own check needs, which keeps a finding attributable to that role. Persona design draws on~\citep{bui2025mixture,liu2026beyond}, and browser-based simulation follows prior usability work~\citep{xiang2024simuser,lu2025uxagent}. In the example of building a to-do application (Figure~\ref{fig:applooper-conversation-example}), a virtual user flags a fixed reminder interval. The owner confirms the requested changes, the Development Agent submits a revision, and retesting confirms that the new Today filter is present and reports which reminder options still need verification.

\subsubsection{Development Agent.}
\leavevmode\label{sec:development-agent}
\hypertarget{role:development-agent}{}The development agent implements each roadmap task and revises the candidate after feedback. Only it can modify product source code, leaving the other roles' checks read-only. Its \hypertarget{module:bounded-development-context}{\textbf{task context}} supplies the current task, frozen requirement, dependencies, confirmed feedback, and previous iteration's state, and the revision stays within that scope. During \hypertarget{module:implementation-revision}{\textbf{implementation and revision}}, the agent fixes defects, adds small features, and improves usability within confirmed constraints in the participant's isolated workspace, choosing conservative, reversible implementations under ambiguity or conflict. Each revision runs relevant tests and visits the real-browser preview and every declared page. Major capability changes await owner confirmation as subsequent tasks, as in the last message of Figure~\ref{fig:applooper-conversation-example}(a). The result is a \hypertarget{module:versioned-development-evidence}{\textbf{candidate version and evidence}}. The version is identified by a hash of the application's source files. Its trial entry point and evidence such as test records and screenshots are recorded under that hash, so the owner tries the version that the evidence describes.

The development agent's role template reads as follows.
\begin{AppLooperPrompt}
You are the only developer agent that may modify source code in the current workspace. Inspect existing code first and build on it.

Strictly obey active guardrails. Bugs, small in-guardrail features, and usability issues from experience feedback may be implemented directly; unapproved major capabilities or large changes to existing capabilities must be deferred.

Ambiguous, conflicting, or low-confidence decisions must keep the current implementation or take the most conservative reversible default. Run relevant tests, prepare a real experience entry, and preserve absolute screenshot paths when possible.
\end{AppLooperPrompt}

Each development task adds the following call instruction.
\begin{AppLooperPrompt}
Complete only current\_task, test it, and return immediately; do not start the next task in task\_graph. Preserve existing behavior for unspecified details or use the most conservative reversible default. Fix bugs, copy, layout, usability, and small features directly. For an interactive product, create or update a runnable Experience Twin, visit every declared view, and record the verification in tests. Each changes item must name an atomic user-visible behavior.
\end{AppLooperPrompt}

\subsubsection{Virtual-User Agent Cohort.}
\leavevmode\label{sec:virtual-user-cohort}
\hypertarget{role:virtual-user-cohort}{}Virtual users inspect each new runnable candidate from multiple target-user perspectives, gathering experience evidence with read-only access to the product implementation. A structured planning call uses the target users, application type, and owner description to request four to eight fictional \hypertarget{module:explicit-personas}{\textbf{personas}} and browser task scripts for desktop and mobile use. Each persona records its covered population, demographics, location and role, technical proficiency, device, motivations, constraints, usage context, habits, and an interaction script. Virtual users perform \hypertarget{module:browser-end-to-end-tasks}{\textbf{real-browser end-to-end tasks}} with synthetic test identities, so the findings come from operating the interface. In the example of building a to-do application (Figure~\ref{fig:applooper-pc-mobile}), an office-worker virtual user first finds the creation entry point, creates a categorized item with a deadline, and checks that both are visible in the list. Login and registration use controlled study verification procedures, synthetic identity data, and communications confined to the study environment. The cohort's \hypertarget{module:candidate-bound-findings}{\textbf{experience findings}} record the issues or interaction failures encountered in the current candidate, together with the pages, controls, and interaction traces involved, so each issue is stored with the version in which it appeared.

Persona generation uses the following call instruction.
\begin{AppLooperPrompt}
Invent 4 to 8 fictional end users for this specific app. Ground every person in the owner's app\_type, audience, and description only. If audience text conflicts with the app type or description, follow the app type and description. Do not copy the participant's real identity or invent the participant's demographics. Each person must be a distinct everyday user of THIS app, with a unique name, role, scenario, motivation, constraints, habits, and a short task\_script of concrete UI actions. Split the cohort across devices: about half primarily use a computer browser (desktop\_web), about half a phone browser (mobile\_web). task\_script actions must be ordinary user actions on this app, not architecture or developer work.
\end{AppLooperPrompt}

Each virtual user receives the following role template.
\begin{AppLooperPrompt}
You are a read-only experience agent and must strictly play the given persona. Do not edit, write, or modify source code.

Complete end-to-end tasks through real Computer Use. Do not fake UI experience with source reading, curl, or API calls.

Record evidence and screenshots item by item, and submit only problems you actually encountered. If you cannot operate the UI, report failure; never invent success.
\end{AppLooperPrompt}
\subsubsection{Testing Agent.}
\leavevmode\label{sec:testing-agent}
\hypertarget{role:testing-agent}{}On each new runnable candidate, the testing agent performs read-only gray-box checks, which operate the running application and also read its source code, looking for technical failures without changing the candidate. It first \hypertarget{module:reproduce-latest-failure}{\textbf{reproduces the latest failure}}, returning to failures recently reported or recovered from by the development agent and using relevant source code to locate causes, before treating a failure as still open. It then \hypertarget{module:existing-regression-tests}{\textbf{runs existing regression tests}}, checking whether the current task, key interfaces, and visible pages still hold in a real browser. Suppose the development agent reports that saving a task with a deadline failed and has since been repaired. The testing agent would first save such a task again in the browser, and then check that the list, the filters, and the other visible pages still work. It reports only \hypertarget{module:observed-defects}{\textbf{observed defects}}, with their evidence and affected candidate versions, to feedback routing and system-side evaluation.

The testing agent's role template reads as follows.
\begin{AppLooperPrompt}
You are AppLooper's read-only developmental test agent. Inspect the current runnable build and never modify product source.

Reproduce the developer's latest reported or recovered failure first, then check the current task, critical interfaces, and visible UI. When the same command already failed, diagnose the root cause before running it again. Exercise runnable web UI through a real browser and record evidence. Report only observed defects; do not treat later unimplemented tasks as defects.
\end{AppLooperPrompt}

\subsubsection{Owner-Intent Simulation Agent.}
\leavevmode\label{sec:owner-intent-agent}
\hypertarget{role:owner-intent-simulation-agent}{}The agent operates in two modes. Feedback precheck examines a newly submitted piece of owner feedback before the development agent acts on it. Post-revision retesting operates the revised candidate and checks whether the behavior that the feedback concerned now matches the owner's statements. In both modes, the material on which the agent bases its judgments is restricted to \hypertarget{module:bounded-owner-evidence}{\textbf{owner statements}}, comprising the frozen requirement, confirmed constraints, and explicit owner feedback. In feedback precheck, the agent checks consistency, conflicts, missing information, and actionable acceptance points solely against the owner's original statements, cited by identifier, without operating the interface or rewriting feedback. Suppose the owner of the to-do application asks for reminders to come earlier. The precheck would find the request consistent with the frozen requirement of reminders before deadlines, note that the message states no interval, and list an earlier reminder as the point for development to check. After revision, the agent conducts \hypertarget{module:targeted-retesting}{\textbf{targeted retesting}} for at most two rounds on relevant scenarios in the real interface, adding evidence from actual interactions. The agent reports attempts, outcomes, and unresolved risks as \hypertarget{module:simulated-owner-evidence}{\textbf{simulated owner-intent evidence}} when evidence is sufficient and otherwise abstains. Both modes produce results with this label, stored separately from actual owner acceptance records, so a simulated check remains distinct from the owner's own decision.

The agent's role template reads as follows.
\begin{AppLooperPrompt}
You are a read-only product-owner intent agent acting as the owner-intent simulation agent, not a target-user persona and not the real owner. You have two strictly separated modes. In feedback\_precheck mode, use only the itemized frozen intent, current verbatim user message, explicit prior messages, confirmed constraints, and source\_id values to check consistency, conflicts, missing information, and executable acceptance checks before feedback reaches development; do not use Computer Use, reject or rewrite real-user feedback, and provide only source-backed developer guidance. In post\_fix\_replay mode, inspect one repaired scenario through real Computer Use.

In either mode, do not infer demographics, personal preferences, or unstated needs; abstain when evidence is insufficient. Do not modify source code, guardrails, acceptance results, release conclusions, or attestations, and do not speak as the owner.
\end{AppLooperPrompt}

Feedback precheck adds the following authority boundary and output contract.
\begin{AppLooperPrompt}
Advisory only. Preserve the real user's message verbatim. Do not reject, rewrite, expand, or silently change it; do not make a release verdict.

Use only supplied source IDs. List concrete acceptance checks for development when explicitly supported.
\end{AppLooperPrompt}

Post-revision retesting adds the following authority boundary.
\begin{AppLooperPrompt}
Advisory only. Do not write guardrails, release verdicts or attestations, evaluator results, and do not speak as the real owner. Use real Computer Use for every task\_script step. Never infer success from source code, API/HTTP responses, translation keys, or a developer claim. A PASS needs a visible interaction plus fresh rendered evidence for that exact step.
\end{AppLooperPrompt}

\section{AppLooper Workflow}
\label{sec:workflow}

This section presents the workflow of Figure~\ref{fig:framework}. Feasibility judgment and lifecycle initialization fix the requirement and plan the work, candidate acceptance and release end with the owner's decision, and between these stages the guardrail management keeps the constraints confirmed by the owner in force during the development that follows while lifecycle orchestration coordinates checks, revisions, and retests on each candidate version.

\subsection{Feasibility Judgment and Initialization}
\label{sec:app_accepted}

Before development begins, Application Feasibility Judgment screens the owner's initial intent to establish a safe, comparable scope for prototype tasks in the study. It applies a \hypertarget{module:deterministic-scope-policy}{deterministic scope policy} followed by online model refinement. Before the two stages, an intent that closely repeats the example given in the interface is rejected with a request for a different idea. The first stage runs without a model, normalizing the initial intent and checking it against fixed keyword lists in a fixed priority order. It rejects credential theft, malware, and security circumvention, and excludes medical diagnosis and treatment, control of real vehicles, weapons and nuclear facilities, and high-risk financial or legal decisions. Out-of-scope goals include artificial general intelligence and training a foundation model from scratch. Requests dependent on real sensitive accounts, production systems, or hardware control are rewritten as safe prototypes that use fictional data and simulated interfaces. Suppose an owner asks for a bookkeeping tool that reads the transactions of a real bank account. The idea is kept, with the added condition that the prototype uses synthetic sample data and simulated interfaces. A hospital diagnostic assistant matches the medical keywords and is rejected, whereas a personal to-do and reminder tool matches no list and proceeds to refinement.

\hypertarget{module:online-model-refinement}{}In the second stage, a single model call re-examines the intent and returns an admit, rewrite, or reject judgment, a brief rationale, and a complete frozen requirement comprising target users, application type, and core problem, which the owner can then confirm. In the to-do example of Figure~\ref{fig:framework}, the frozen requirement names busy people as the target users, a simple to-do application as the type, and organizing tasks, setting due times, and getting reminders as the core problem. Unavailability, timeout, or invalid model output falls back to the first-stage result. Online refinement processes only retained requests and uses a prompt template matching the interface language. \hypertarget{module:owner-confirmed-frozen-requirement}{}Owner confirmation establishes the frozen requirement as the shared initial baseline. Later explicit feedback and owner-approved constraint changes are recorded separately, each together with the candidate versions and acceptance evidence it concerns. A later request for a Today filter, for instance, is stored as feedback on the candidate version on which it was made, and the frozen requirement stays as confirmed.

Online model refinement uses the following instruction, with the initial intent appended as JSON.
\begin{AppLooperPrompt}
You are a strict admission gate for a software-prototype user study. The decision field must be one of PASS, REPHRASE, or BLOCK. Ordinary and safe software ideas should PASS. Only low-difficulty and non-serious applications are allowed. Medical diagnosis/treatment, autonomous driving or vehicle control, weapons and nuclear facilities, high-risk financial or legal decision-making, and clearly high-difficulty goals such as AGI or training a foundation model from scratch must be BLOCK. If the request depends on unavailable hardware, real sensitive accounts, or production systems but a safe simulated prototype is feasible, use REPHRASE. If it is clearly malicious, illegal, credential-stealing, or cannot be safely rewritten, use BLOCK. The gate is not a success-probability predictor.
\end{AppLooperPrompt}

After owner confirmation, lifecycle initialization produces the \hypertarget{module:requirement-roadmap}{Requirement Roadmap}, decomposing the frozen requirement into confirmed constraints, dependent tasks, and virtual-user personas that later checks and the owner's inspection can share. Each constraint contains only content the owner stated or approved, acceptance conditions that can be verified, and the exact passage of the development intent that supports it. Tasks record dependencies, associated constraints, verification methods, and status. Every participant works in a separate, versioned workspace with the same data and AI capabilities available. Each participant account can reach only its own workspace, and application previews are shown in sandboxed iframes that cannot access the surrounding page.

Planning and orchestration calls use the following role template.
\begin{AppLooperPrompt}
You are a read-only workflow orchestration agent. Manage dependencies, state, personas, guardrails, and acceptance; never modify source code.

Only content the user has explicitly stated or approved may become a guardrail; inferences cannot become guardrails. For ambiguity, conflict, or low confidence, keep the current workspace behavior or the most conservative reversible default; do not expand it into a guardrail or ask the user to confirm.

Personas must be detailed and cover every declared target-user segment.

During acceptance, independently inspect code and evidence and actually run regression tests; do not trust the developer's verbal claims.
\end{AppLooperPrompt}

The Requirement Roadmap call adds the following instruction.
\begin{AppLooperPrompt}
Plan or re-plan the workflow from all active guardrails, candidate requirements, and the existing workspace implementation. Active guardrails are confirmed, non-negotiable constraints; candidate requirements, observations, and discussion are inputs or verification evidence and must never be promoted to guardrails automatically. Inspect the existing code first. Preserve unspecified behavior and configuration; where none exists, choose the most conservative, reversible option with the smallest new capability. Split only long-lived boundaries explicitly stated in intent into guardrails; every guardrail needs text, acceptance criteria, and a verbatim source\_quote. Never turn tester feedback, a repair request, or a one-off task into a guardrail. Every task\_script action must directly exercise an active guardrail, a frozen candidate requirement, or the shared CUJ; do not add feature exploration (for example statistics, streaks, or social features) unless the frozen scope explicitly requires it. Tasks must form an executable DAG.
\end{AppLooperPrompt}
\subsection{Guardrail Management}
\label{sec:guardrails}

The guardrail management keeps every revision and check within what the owner has confirmed, from initialization to release. A guardrail is a confirmed constraint of the Requirement Roadmap. In the example of building a to-do application (Figure~\ref{fig:framework}), issuing reminders before deadlines is one guardrail, and every later candidate must satisfy it. \hypertarget{module:guardrail-admission}{}The system admits a constraint proposed by the roadmap as a guardrail when its supporting passage appears word for word in the owner's development intent. \hypertarget{module:owner-statement-classification}{}During development, the statements in each owner message are classified as observations, discussion, candidate requirements, or confirmed constraints, and a confirmed constraint can add, replace, or disable a guardrail. The change takes effect when the classification confidence is at least 0.8, the supporting passage appears word for word in that message, and the passage is worded as a lasting rule, with terms such as \emph{must}, \emph{never}, or \emph{always}, or answers a confirmation request from the system. A statement that every task must always show its deadline can thus add a guardrail, whereas a remark that the list looks crowded is recorded as an observation. In other cases the message is kept on record and development continues under the current guardrails. Findings from virtual users and the testing agent prompt revisions, while the guardrail set changes through the owner's statements. Each change is appended to a log with a revision number, which preserves the earlier states of the set. \hypertarget{module:guardrail-integrity-check}{}Agents receive the guardrails as input. The system compares a short fixed summary of the active set before and after each agent call and halts the workflow if an agent has altered it. System-side evaluation reports a result with evidence for every active guardrail (Section~\ref{sec:system-candidate-threshold}), and a candidate evaluated under an earlier set is evaluated again once the set changes.

\subsection{Orchestration and State Management}
\label{sec:lifecycle-orchestration}

Orchestration coordinates the agents through three components, which route feedback, track tasks and their state, and schedule roles. At each stage, it supplies the relevant prompts, frozen requirement, current candidate version, task context, and output constraints. Chinese or English prompt templates follow the interface language.

\subsubsection{Feedback Routing and Group Management.}
\leavevmode\label{sec:feedback-clusters}
After virtual-user and testing agents check a candidate, feedback routing takes the issues they report, together with the supporting evidence, and decides which are revised and retested together. Routing groups issues that share an interface target or, when none is recorded, that have similar descriptions of root causes and affected behaviors. In the example of building a to-do application (Figure~\ref{fig:framework}), there can be a case in which one virtual user reports that a deadline cannot be chosen on a phone and another that a saved task shows no deadline. Both concern the deadline control and form one group. Each group of related issues lists the agents that reported it, an evidence summary, a priority, and its candidate version, and it is the unit for revision and targeted retesting. For the deadline group, these are the two virtual users, what each saw, the priority of the more severe report, and the version both tried. Owner feedback follows the separate route described below and, like owner-intent evidence, enters state management as a separate record marked as owner-written or simulated.

Owner issues submitted in the isolated trial window retain their original-text records and proceed directly to development revision through a separate route that keeps the owner's wording distinct from agent-written findings. Their revision is followed by targeted retesting and owner inspection. Before an owner's feedback enters development, feedback precheck compares it with the requirement and constraints the owner has confirmed and with the owner's earlier messages. Its output is recorded as simulated owner-intent evidence alongside the original feedback. The records of a candidate version therefore show separately what the agents found, what the owner requested, and how the agent assessed that request. In this to-do example, if the owner asks for reminders to come earlier, the records would hold the virtual users' findings on that candidate, the owner's sentence as written, and the precheck's note that the message states no interval.

\subsubsection{Task Dependency and Lifecycle State Management.}
\leavevmode\label{sec:task-lifecycle-state}
This component stores the roadmap, the grouped issues, and owner feedback in one place and keeps task dependencies, candidate versions, test results, and feedback status up to date, so the next check starts from the current state. The roadmap displays the states not started, in progress, implemented pending verification, verified, blocked, or deferred, with supporting evidence and next steps. In the example of building a to-do application (Figure~\ref{fig:framework}), the requirement covers recording and categorizing items, setting deadlines, and issuing reminders before deadlines. These become three confirmed constraints with dependent tasks for creation and categorization, deadline and reminder settings, and core-path verification.

\subsubsection{Role Scheduling and Scenario Management.}
\begin{figure*}[!t]
  \centering
  \includegraphics[height=.65\linewidth,trim=392bp 0 421bp 0,clip,alt={Desktop isolated trial window for the candidate application, with a control for submitting issues found during use.}]{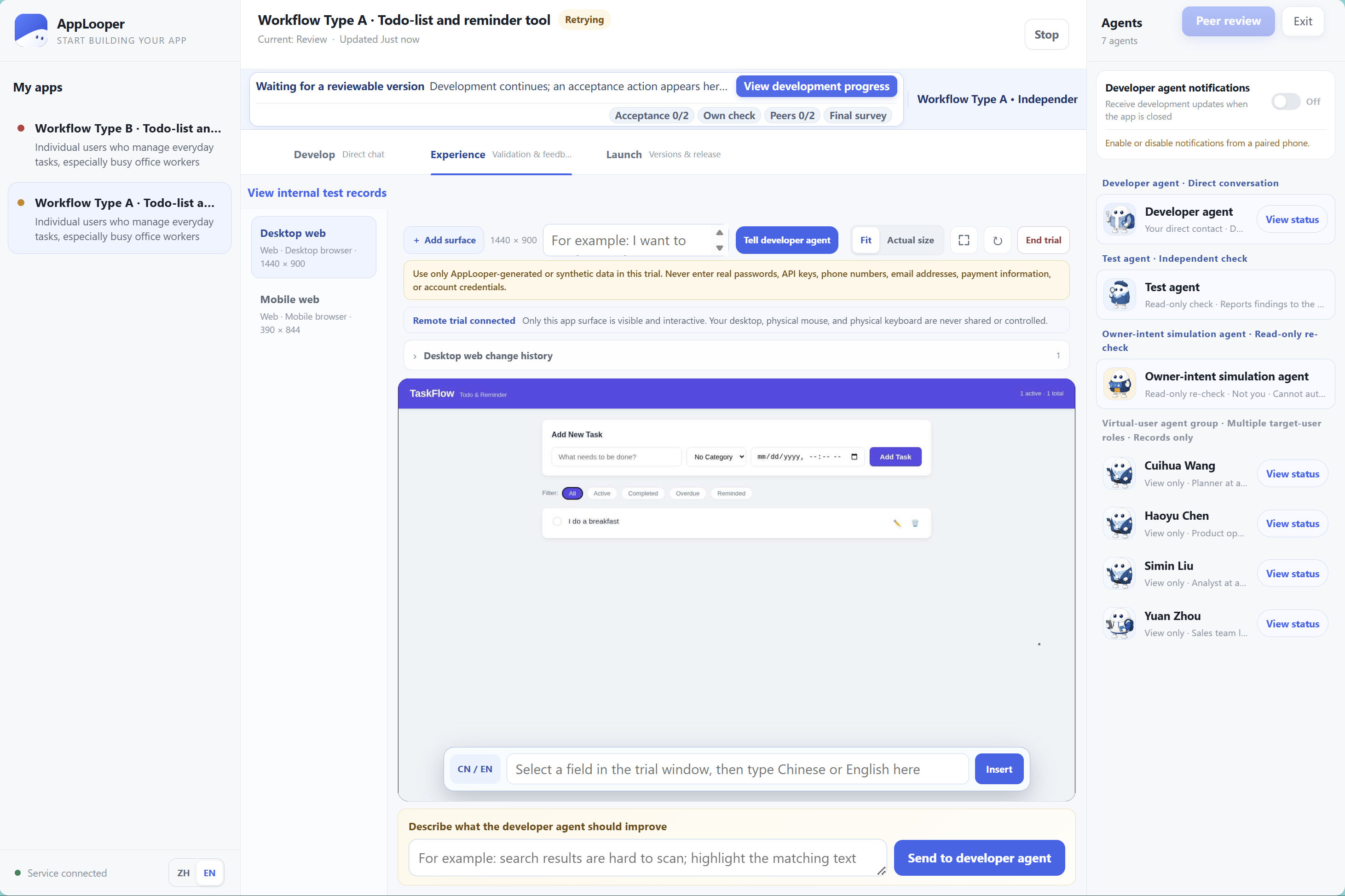}\hfill
  \includegraphics[height=.65\linewidth,alt={Mobile development conversation showing agent testing feedback and revision updates.}]{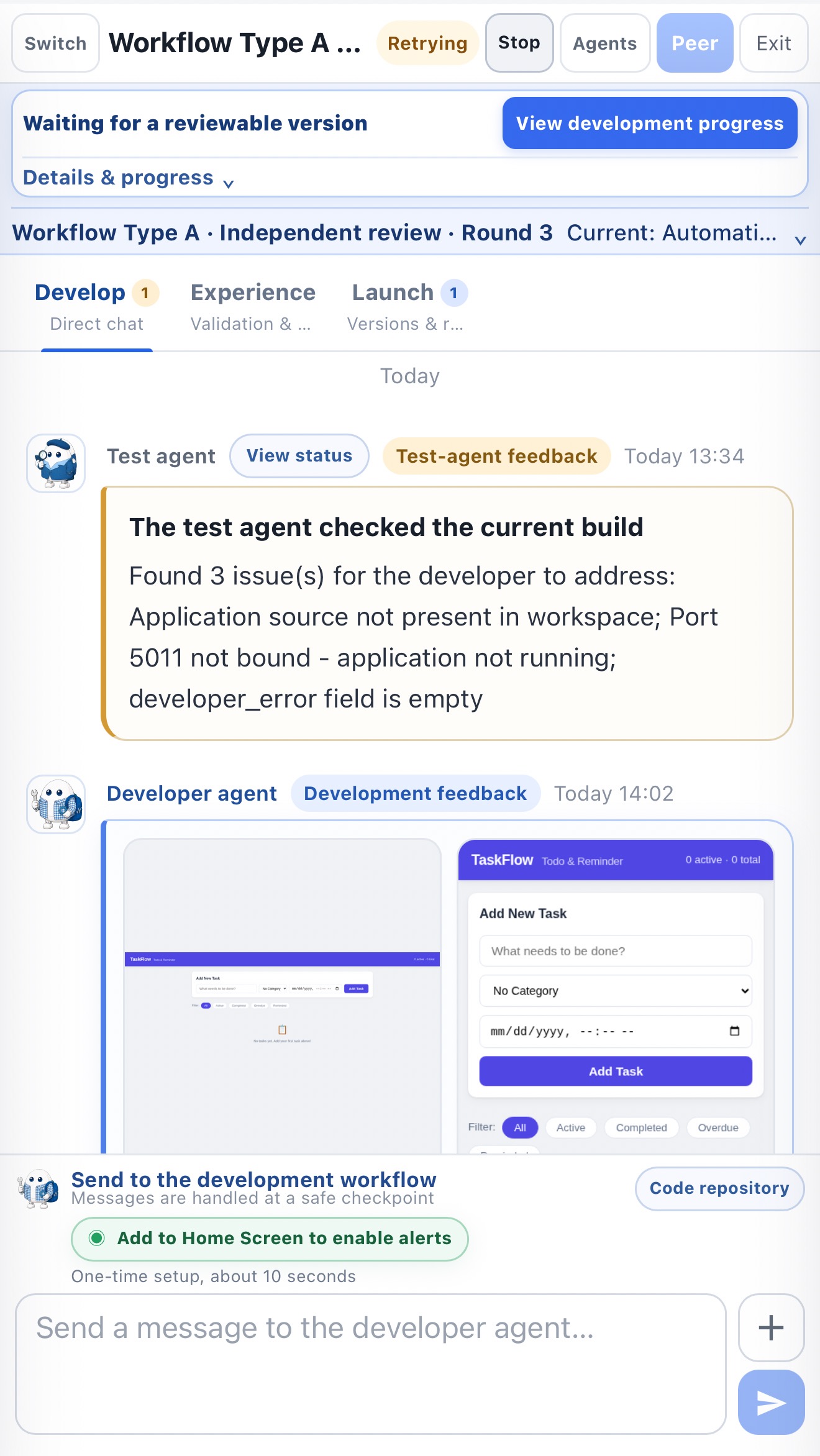}
  \caption{AppLooper's desktop trial and feedback controls (left) and mobile development dialogue (right).}
  \label{fig:applooper-pc-mobile}
  \Description{Two AppLooper interface screenshots. The left shows a desktop window in which the application owner can try a candidate application and report issues. The right shows the development dialogue, testing feedback, and revision updates on a mobile phone.}
\end{figure*}

\leavevmode\label{sec:role-scheduling}
This component selects the agent for the current checkpoint and writes a targeted retesting scenario for the task or feedback at hand, so the follow-up check addresses the behavior just changed. Each group of related issues becomes a scenario for one candidate version, with initial conditions, interaction steps, expected and actual outcomes, owner judgment, temporary-state cleanup, and the revision and kind of check behind the evidence. Scenarios execute in the order open $\rightarrow$ interact $\rightarrow$ judge $\rightarrow$ clean up. When the candidate changes, the conclusions reached for existing scenarios must be validated again. Critical user journeys are generated from the frozen requirement. In the example of building a to-do application (Figure~\ref{fig:applooper-pc-mobile}), a target user in a fresh session finds the main entry point, creates and categorizes an item, sets a deadline, and confirms a visible result without blocking errors. After revision, state management schedules the project's regression tests and real-browser interaction checks, alongside relevant targeted scenarios. Their results update the candidate's state and evidence, along with the grouped issues, tasks, and the roadmap. This evidence-driven iteration follows the feedback approaches of \citep{madaan2023self} and \citep{shinn2023reflexion}. In this example, scenario management turns a virtual user's finding that the deadline is missing after creating an item into a targeted scenario that creates an item with a category and deadline and expects the deadline in both list and detail views. Role scheduling then assigns the virtual user who reported the issue to repeat it on the revised candidate. At the inspection checkpoint, the system presents the scenario with the role that found the issue, the evidence, the version's changes, and the interface target, and guides the owner through the same steps, so the owner's check is made on the revised candidate. These version-specific explanations align with \citep{yan2024ivie} and the natural-language steps, interface-element mapping, and stepwise verification of mixed-initiative web automation \citep{chen2023miwa}.

\subsection{Candidate Acceptance and Release}
\label{sec:candidate-acceptance}

Release proceeds through the System-Side Candidate Threshold and Owner Acceptance, first asking whether autonomous checks are sufficient to pause iteration, then leaving the release decision with the owner. 

\subsubsection{System-Side Candidate Threshold.}
\leavevmode\label{sec:system-candidate-threshold}
System-side evaluation judges whether iteration should pause for the owner's inspection. It is a separate call that does not share the development agent's conversation and reads frozen constraints, layered test results, critical user journey evidence, the findings recorded for the candidate, and defect states. It reports constraint satisfaction, regression results, issues, screenshots, and completed tasks. Issues are graded low, medium, high, critical, or blocking. Any reported issue returns the candidate to revision, with critical and blocking issues addressed first. A reminder that does not appear before a deadline, for example, would be reported under the reminder constraint and return the candidate to revision. Shared tests are executed, and critical journeys run in real browsers with retained interaction evidence.

System-side evaluation uses the orchestration role template (Section~\ref{sec:app_accepted}) with the following call instruction and addition.
\begin{AppLooperPrompt}
Blind standardized release evaluation. Return exactly one guardrails row for every frozen\_requirements item.

This turn is independent candidate review, not planning. Use the minimum read-only product-file inspection and Bash tests needed for evidence. Never inspect \textup{[internal directory name]}, agent/session metadata, conversation or progress logs, prior verdicts, screenshots from earlier agents, workflow labels, or condition identifiers. Do not infer which study workflow produced the artifact.
\end{AppLooperPrompt}

\subsubsection{Candidate Evaluation.}
\leavevmode\label{sec:candidate-evaluation}
The first qualifying candidate, or the current candidate once a preset round limit is reached, is frozen as the \hypertarget{module:immutable-acceptance-version}{acceptance candidate}, and autonomous iteration stops, so the version the owner inspects stays fixed until the owner acts. The system threshold and the round limit determine which version the owner inspects, and the release decision remains with the application owner. When a defect of the highest priority blocks a core task, repair, regression testing, browser retesting, and acceptance start again. Every revision creates a new candidate requiring complete evaluation, and prior approvals become invalid. If the Today filter is changed after the owner has inspected a candidate, for example, the result is a new candidate that is evaluated and inspected again. The evidence from testing, feedback, revision, retesting, and evaluation is kept with the version it concerns. The owner's inspection and approval therefore concern one specific candidate and its evidence.

\subsubsection{Owner Acceptance.}
\leavevmode\label{sec:owner-acceptance}
The owner acceptance checklist brings together frozen requirements, critical journeys, current tests, unresolved high-priority issues, reviewed and pending feedback, labels marking which evidence is simulated, and known limitations for the release decision. The interface organizes key feedback as the issue, the evidence, the change, and the inspection target, with expandable full evidence. For the Today filter request of the to-do example, these would be the owner's request, the retest screenshot, the added filter, and the Today control to open in the trial window. Because supervision interfaces may influence inspection and confidence \citep{grunde2026overseeing}, authorization includes owner use of the isolated trial window (Figure~\ref{fig:applooper-pc-mobile}, left). Issues found there can return directly to development, and the system records their candidate and interface target. After completing the core tasks, the owner approves, returns, or defers the candidate. The system saves the owner's operations, inspection results, and final decision as the owner acceptance record and stores it separately from the results of virtual-user testing and simulated owner-intent retesting.

\subsubsection{Release Record.}
\leavevmode\label{sec:accountable-release-record}
\hypertarget{module:human-authorized-final-release}{}The owner's release decision generates a release record that brings together the frozen requirement, candidate version, system evaluation, evidence references, owner trial record, and final decision. The results of virtual-user testing and owner-intent retesting remain labeled as simulated evidence, and the owner's own actions and later operational data are recorded apart from them. The record supports tracing the approved version, its requirements and evidence, the feedback items revised and retested, and the basis of the final inspection. For this to-do application example, the record might name the approved version, the frozen requirement of reminders before deadlines, the system evaluation of that version, the Today filter request with its revision and retest, the owner's trial of the Today control, and the owner's approval.

\section{User Study}
\label{sec:user-study}

We evaluated AppLooper through an online within-participant paired user study against Claude Code Loop, a baseline without AppLooper's virtual-user, testing, and owner-intent simulation agents. Participants developed applications from their own ideas and evaluated both the applications and the development experience through the release decision. The study examined owners' assessments of application quality, effort, understanding, and self-reported trust at release. It also explored how owners participated in development, interpreted agent feedback and revisions, and experienced waiting and returning to ongoing work. Cumulative LLM input and output volume provided resource-use context for the comparison.

\subsection{Study Setup}
\label{sec:study-participants-setup}

The study setup describes the participants and the configuration of the AppLooper and Claude Code Loop conditions. Each participant used both conditions, and the shared settings and condition differences are given in Table~\ref{tab:study-settings}.

\subsubsection{Participants.}
\label{sec:study-participants}

We recruited through campus platforms and referrals via social messaging. Any adult aged 18 or older could enroll, irrespective of programming background. Questionnaire records were available for 33 participants, of whom 30 were included in the paired analysis. The other three were not included because they declared their own judgments to be biased or because the information conveyed in their interactions with the study organizers was clearly insufficient to treat them as valid participants. Reported age bands spanned 18 to 62 years. Ten selected a computing or software student or professional role. Among the remaining 20, six selected non-computing student, one selected research or academia, one selected product or design, operations or sales, and \emph{other}, and twelve selected only \emph{other}. The multiple-choice item provided preset categories, including an unspecified \emph{other} option. Mean self-rated programming and AI-tool experience were 1.50 and 2.07 on scales from 0 to 5 (medians 1 and 1.5).

\subsubsection{Conditions and Configuration.}
\leavevmode\label{sec:study-configuration}
Both development agents were built on Claude Code, with Agnes AI's Agnes~2.5-flash as the inference engine. The same model supported the virtual-user, testing, and owner-intent roles. Feasibility judgment and the shared structured evaluator used Agnes~2.5-flash at temperature~0. The conditions shared initial requirements, core tasks, evaluation criteria, and a 20-round cap (Table~\ref{tab:study-settings}). Available materials were owner uploads, retrieved reference material saved with its origin, explicitly labeled synthetic test data, and owner-configured inference services. Secret values were supplied through controlled configuration. Each participant account ran in a separate unprivileged container, with its own internal network and persistent workspace. Containers used read-only root filesystems, dropped operating-system capabilities, prohibited privilege escalation, and were limited to at most 3 CPU cores, 6~GB of memory, and 1,024 processes. External access passed through a controlled allowlist proxy.

To examine the gains that AppLooper's multi-role checks and version-specific evidence records bring to owners' assessments of application quality, effort, understanding, and release trust, we used Claude Code Loop as the control baseline. It combined the Claude Code model and tool loop~\citep{liu2026dive} with an external evaluation, feedback, and redevelopment loop based on the evaluator and optimizer pattern~\citep{anthropic2024effectiveagents}. Its single development agent handled implementation, self-checking, and participant feedback, and revised from the findings of the shared independent candidate evaluation. Compared with AppLooper, it removed virtual-user simulation, independent testing, feedback prechecks, and post-revision retesting, and the display of which revisions and review followed each feedback item.

\begin{table*}[tbp]
  \centering
  \caption{Core study settings for AppLooper and the control baseline. The upper portion summarizes shared settings, and the lower portion compares differences.}
  \label{tab:study-settings}
  \fontsize{9}{11}\selectfont
  \setlength{\tabcolsep}{4pt}
  \setlength{\arrayrulewidth}{0.4pt}
  \renewcommand{\arraystretch}{1.1}
  \begin{tabularx}{\textwidth}{@{}>{\raggedright\arraybackslash}p{.15\textwidth}>{\raggedright\arraybackslash}X>{\raggedright\arraybackslash}X@{}}
    \toprule
    \multicolumn{3}{@{}l}{\textbf{Shared settings}} \\
    \addlinespace[4pt]
    Model and task & \multicolumn{2}{>{\raggedright\arraybackslash}p{\dimexpr.85\textwidth-2\tabcolsep\relax}@{}}{Same coding model, initial task requirements, evaluation criteria, and core usage tasks.} \\
    \addlinespace[3pt]
    Independent candidate evaluation & \multicolumn{2}{>{\raggedright\arraybackslash}p{\dimexpr.85\textwidth-2\tabcolsep\relax}@{}}{Shared independent evaluation checks candidate versions against frozen requirements, critical user journeys, and layered tests.} \\
    \addlinespace[3pt]
    Trials and feedback & \multicolumn{2}{>{\raggedright\arraybackslash}p{\dimexpr.85\textwidth-2\tabcolsep\relax}@{}}{Trials begin with the first runnable version. Feedback can be submitted through the trial interface and development conversation.} \\
    \addlinespace[3pt]
    Iteration cap & \multicolumn{2}{>{\raggedright\arraybackslash}p{\dimexpr.85\textwidth-2\tabcolsep\relax}@{}}{At most 20 rounds. The baseline count restarts when a delivered candidate is reopened for revision. At the cap, the current result is frozen and submitted for participant acceptance with completed checks and unresolved issues.} \\
    \addlinespace[3pt]
    Release authorization & \multicolumn{2}{>{\raggedright\arraybackslash}p{\dimexpr.85\textwidth-2\tabcolsep\relax}@{}}{Participants operate the application and approve it or return it for revision. Any modification creates a new version awaiting acceptance and invalidates approval of the previous version.} \\
    \arrayrulecolor{black!28}
    \specialrule{0.45pt}{5pt}{5pt}
    \arrayrulecolor{black}
    \textbf{Differences} & \textbf{AppLooper} & \textbf{Control baseline} \\
    \addlinespace[4pt]
    Roles and feedback activities & Development-agent iteration with target-user simulation, independent testing, owner-intent prechecks, and post-revision retesting, showing which revisions and review followed each piece of feedback. & The development agent implements, self-checks, handles participant feedback, and revises based on external evaluation feedback. \\
    \addlinespace[5pt]
    Early readiness decision & Passes the shared evaluation and the current round's virtual-user and independent testing checks, and owner-intent retesting yields an acceptable result when applicable. & The development agent reports readiness and the shared independent candidate evaluation passes. \\
    \bottomrule
  \end{tabularx}
\end{table*}

\subsection{Study Procedure}
\label{sec:study-procedure-analysis}

The procedure had three parts. Participants stated an application idea that both workflows then developed, tried the candidate versions and made a release decision in each workflow, and finally answered two questionnaires. Steps that also belong to the workflow of Section~\ref{sec:workflow} are described briefly here, with the subsection that gives the details cited in parentheses.

\subsubsection{Requirement Elicitation and Condition Assignment.}
\label{sec:study-elicitation}

Participants read and confirmed informed consent, supplied background information, and specified target users, application type, and core problem. The initial prompts asked who the app was for, what type of app they wanted, and what problem it should solve. Participants chose their own ideas to ground trials and feedback in goals they cared about. Feasibility screening then rejected or rewrote ideas outside a safe prototype scope and returned a frozen requirement stating target users, application type, and core problem that the participant could confirm (see Section~\ref{sec:app_accepted} for details). Each idea thereby supplied the same initial requirements to both workflows. Each participant experienced both conditions, producing 60 workflows.

Both workflows were initialized together and could progress asynchronously. The median span from workflow creation to a participant's last release judgment was 1.36 days (range 0.15 to 9.24). Condition order was randomized across participants. For each participant, the interface and questionnaires labeled the two workflows A and B, with the assignment of conditions to letters randomized and then kept fixed. The questionnaires identified workflow types by agent configuration.

\subsubsection{Development, Inspection, and Acceptance.}
\label{sec:study-development-acceptance}

\begin{figure}[tbp]
  \centering
  \includegraphics[width=\linewidth,trim=12.75bp 5.25bp 13.5bp 12.75bp,clip,alt={Development progress dashboard showing the current stage, interaction-round count, and core requirement statuses.}]{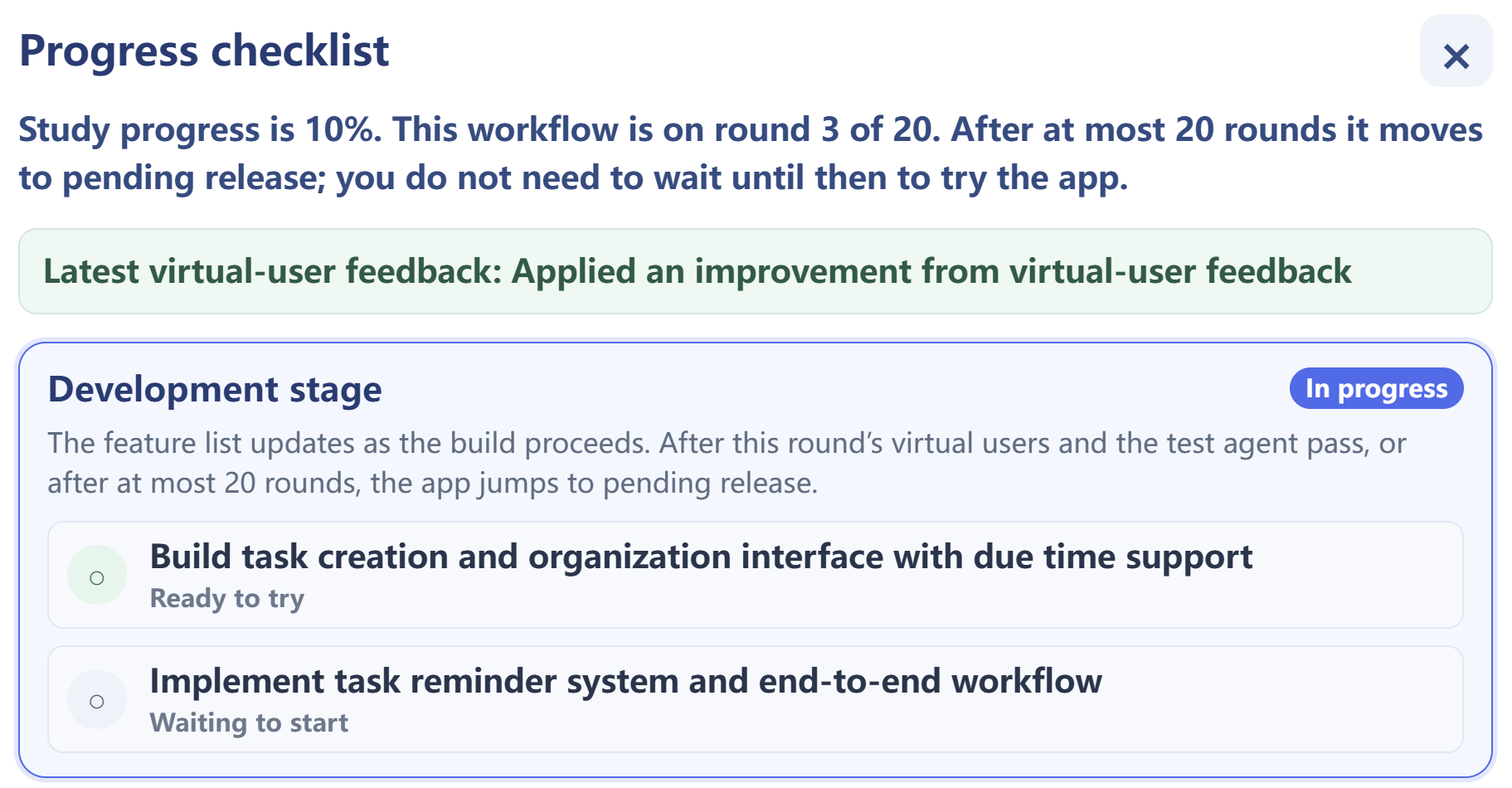}
  \caption{AppLooper's progress checklist shows the current stage, development round, and status of core requirements.}
  \label{fig:applooper-view-progress}
  \Description{Screenshot of the development progress dashboard. The top shows the current stage and interaction rounds. The bottom lists the application's core requirements and the status of each.}
\end{figure}

Both conditions shared the trial, feedback, and progress interfaces. Participants could try the first runnable version, report issues from the trial interface or development conversation (Figure~\ref{fig:applooper-pc-mobile}), and inspect later revisions. The progress checklist showed the current stage, rounds, and requirement status. AppLooper also showed virtual-user feedback and associated changes when available (banner in Figure~\ref{fig:applooper-view-progress}).

Each condition entered candidate acceptance when readiness criteria were met or the 20-round cap was reached. At that point autonomous iteration stopped and one version was frozen for the participant's release decision (see Section~\ref{sec:candidate-acceptance} for details). In the baseline, the development agent reported readiness, followed by the shared independent candidate evaluation, a separate call that checks the candidate against the frozen requirement, critical user journeys, and layered tests (see Section~\ref{sec:system-candidate-threshold} for details), and passing moved the version to pending release. AppLooper additionally required the current round's virtual-user and independent testing checks to pass and, when applicable, owner-intent retesting to yield an acceptable result. At the cap, or when repeated errors stopped automatic progress, both conditions froze the current result and presented it with completed checks and unresolved issues, as did the baseline after three consecutive rounds without new product changes. The interface highlighted the release control and generated a pre-release checklist of the application's core functions, each with brief steps for verifying it in the trial window. Participants had to try each listed function in the application and mark it as passed or failed before approving the version or returning it for revision. Any change created a new version awaiting acceptance and voided prior approval (see Sections~\ref{sec:candidate-evaluation} and~\ref{sec:owner-acceptance} for details).

\subsubsection{Measures and Analysis.}
\label{sec:study-measures-analysis}

Participants completed two questionnaires after judging both applications. Five-point ratings covered application quality and usability, process burden, participation, and workflow experience. Direct choices and categorical items covered preferences, trust, and specific experiences. Open questions elicited reasons and incidents. Tables~\ref{tab:study-instrument} and~\ref{tab:study-experience-instrument} give the items, response formats, and outcomes. The [A/B] notation indicates that an item was presented separately for each workflow. The interface displayed each participant's assigned agent configuration alongside A or B. Agreement used 1 (strongly disagree) to 5 (strongly agree), quality used 1 (very poor) to 5 (very good), and burden used 1 (very low burden, almost no effort) to 5 (very high burden, considerable effort). The items drew on work on overall effort~\citep{sauro2009comparison}, participation~\citep{barki1994measuring}, context recovery~\citep{endsley1995toward,endsley1995out}, workload~\citep{hart1988development}, and critical incidents~\citep{flanagan1954critical}. Release trust used a single comparative item.

Direct comparisons used two-sided exact binomial tests on clear preferences. The 15 scale dimensions used paired two-sided Wilcoxon signed-rank tests~\citep{wilcoxon1945individual}, with tie-corrected and continuity-corrected normal approximations and zero differences excluded. We report means, paired mean differences, and matched-pairs rank-biserial correlations $r_{\mathrm{rb}}$~\citep{kerby2014simple}. Participant-level paired bootstrap resampling (50,000 samples) yielded pointwise percentile 95\% confidence intervals without multiplicity adjustment~\citep{efron2000introduction}. The four direct comparisons and 15 scale dimensions formed separate Holm families~\citep{holm1979simple}. Raw $p<.05$ and adjusted $p_H<.05$ indicate item-level and within-family significance. Spearman correlations~\citep{spearman1961proof} related programming experience, AI tool experience, and technical identity to preference, quality difference, finished-application effort difference, and final trust, and 50,000 two-sided permutations~\citep{phipson2016permutation} yielded $p$ values, Benjamini--Hochberg corrected across 12 associations~\citep{benjamini1995controlling}.

We summarized feedback and usage descriptively. Keyword rules assigned each message to the first matching category. Unmatched messages formed a separate category. For open responses, ChatGPT assisted an exploratory coding of their meaning, given the question each response answered and the condition to which its A or B label referred. Examples attributed to one condition were taken from responses that clearly named the workflow they referred to. Case interpretations drew on original feedback, the acceptance records of the corresponding version, and owner decisions. Quotations are presented in English translation.

\begin{table*}[tbp]
\small
\setlength{\tabcolsep}{3pt}
\caption{Application comparison questionnaire items. Items marked 1 to 5 were rated on a five-point scale with labeled ends, from 1 (very poor) to 5 (very good) for Q2, from 1 (very low burden, almost no effort) to 5 (very high burden, considerable effort) for Q13 to Q16, and from 1 (strongly disagree) to 5 (strongly agree) for the others.}
\label{tab:study-instrument}
\begin{tabular}{@{}l>{\raggedright\arraybackslash}p{.62\textwidth}>{\raggedright\arraybackslash}p{.24\textwidth}@{}}
\toprule
ID & Question or statement & Response format or outcome \\
\midrule
\multicolumn{3}{@{}l}{\textbf{Application comparison}} \\ \addlinespace[2pt]
Q1 & Which finished app is better? & Choice, overall application preference (Table~\ref{tab:categorical-owner-results}) \\ \addlinespace[2pt]
Q2 & Quality of the finished app from Workflow [A/B] & 1 to 5, final application quality \\ \addlinespace[2pt]
Q3 & Overall, the finished app from Workflow [A/B] is easier to use & 1 to 5, effort saved when using the finished application \\ \addlinespace[2pt]
Q4 & The main things in the finished app from Workflow [A/B] are easy to finish & 1 to 5, core-task usability \\ \addlinespace[2pt]
Q5 & The finished app from Workflow [A/B] solves the problem this app was meant to solve & 1 to 5, core-problem resolution \\ \addlinespace[2pt]
Q6 & Does the finished app from Workflow [A/B] clearly get stuck, break, or stop you from continuing? & Choice (Table~\ref{tab:categorical-conditions}) \\ \addlinespace[2pt]
Q7 & If you used a similar app again, would you rather use the finished app from Workflow A or Workflow B? & Choice (Table~\ref{tab:categorical-comparisons}) \\ \addlinespace[2pt]
Q8 & Please explain why you chose that answer for the previous question. Use one thing that actually happened. & Open response \\ \addlinespace[2pt]
Q9 & When you compared Workflow A and Workflow B, was there a specific thing that made you suddenly feel which of Workflow A / Workflow B was easier or harder to use? & Open response \\ \addlinespace[2pt]
Q10 & Overall, Workflow [A/B] made the task easier for me & 1 to 5, overall effort saved \\ \addlinespace[2pt]
Q11 & Which workflow felt easier to understand? & Choice, ease of understanding (Table~\ref{tab:categorical-owner-results}) \\ \addlinespace[2pt]
Q12 & From your experience using this system, describe why you chose that answer for question 7. & Open response \\ \addlinespace[2pt]
Q13 & In Workflow [A/B], looking at the progress, chat messages, and notes on the page to figure out what to check felt tiring & 1 to 5, reading and selection burden \\ \addlinespace[2pt]
Q14 & In Workflow [A/B], coming back later and figuring out where things stand and what still needs checking felt tiring & 1 to 5, context-recovery burden \\ \addlinespace[2pt]
Q15 & In Workflow [A/B], waiting for the app to finish, or keeping refreshing to see if it is done, felt tiring & 1 to 5, waiting and monitoring burden \\ \addlinespace[2pt]
Q16 & In Workflow [A/B], opening the app yourself, trying it, and deciding whether it can be released felt tiring & 1 to 5, verification and decision burden \\ \addlinespace[2pt]
Q17 & In Workflow [A/B], which step felt the most tiring? & Choice (Table~\ref{tab:categorical-conditions}) \\ \addlinespace[2pt]
Q18 & How strong did your sense of participation feel in Workflow [A/B]? & Choice (Table~\ref{tab:categorical-conditions}) \\ \addlinespace[2pt]
Q19 & What mainly made you feel that way about participation? Name Workflow A or Workflow B. & Open response \\ \addlinespace[2pt]
Q20 & Workflow [A/B]: after I spoke or chose release / do-not-release, I could see the app change because of that & 1 to 5, perceived influence on changes \\ \addlinespace[2pt]
Q21 & Workflow [A/B]: at times so much changed at once that I could not follow it or judge it & 1 to 5, difficulty keeping up with changes \\ \addlinespace[2pt]
Q22 & When you made the release decision, which workflow's finished app did you trust more? & Choice, self-reported trust at release (Table~\ref{tab:categorical-owner-results}) \\ \addlinespace[2pt]
Q23 & What specific thing strengthened or weakened that trust? Name Workflow A or Workflow B. & Open response \\ \addlinespace[2pt]
Q24 & If you built a similar app again, which workflow would you rather use? & Choice, future workflow preference (Table~\ref{tab:categorical-owner-results}) \\ \addlinespace[2pt]
Q25 & Please explain why you chose that answer for the previous question. Use one thing that actually happened. & Open response \\ \addlinespace[2pt]
\bottomrule
\end{tabular}
\end{table*}

\begin{table*}[tbp]
\small
\setlength{\tabcolsep}{3pt}
\caption{Development and release experience questionnaire items. Items marked 1 to 5 were rated on a five-point scale with labeled ends, from 1 (strongly disagree) to 5 (strongly agree).}
\label{tab:study-experience-instrument}
\begin{tabular}{@{}l>{\raggedright\arraybackslash}p{.62\textwidth}>{\raggedright\arraybackslash}p{.24\textwidth}@{}}
\toprule
ID & Question or statement & Response format or outcome \\
\midrule
E1 & Workflow [A/B]: from building to release, the process felt workable and you would do it again & 1 to 5, overall satisfaction \\ \addlinespace[2pt]
E2 & Which workflow felt more efficient for development? & Choice (Table~\ref{tab:categorical-comparisons}) \\ \addlinespace[2pt]
E3 & Why did that one feel more efficient, or about the same? Name Workflow A or Workflow B. & Open response \\ \addlinespace[2pt]
E4 & Workflow [A/B]: looking back, are you still clear and confident about the app and version you released or did not release & 1 to 5, traceability \\ \addlinespace[2pt]
E5 & Which workflow made it clearer what was happening? & Choice (Table~\ref{tab:categorical-comparisons}) \\ \addlinespace[2pt]
E6 & Was there a moment when you suddenly felt behind, or suddenly understood? Was it Workflow A or Workflow B, and what was happening? & Open response \\ \addlinespace[2pt]
E7 & Workflow [A/B] felt easy for me & 1 to 5, ease of the workflow \\ \addlinespace[2pt]
E8 & Which workflow made you more impatient or anxious? & Choice (Table~\ref{tab:categorical-comparisons}) \\ \addlinespace[2pt]
E9 & When you felt impatient or anxious, what were you thinking? Name Workflow A or Workflow B. & Open response \\ \addlinespace[2pt]
E10 & In Workflow [A/B], when you reached the release decision, do you feel you hesitated? & Choice (Table~\ref{tab:categorical-conditions}) \\ \addlinespace[2pt]
E11 & In Workflow [A/B], several agents interacting made you & Choice (Table~\ref{tab:categorical-conditions}) \\ \addlinespace[2pt]
E12 & During Workflow [A/B] development, what kind of information did you most want to be told clearly? & Choice (Table~\ref{tab:categorical-conditions}) \\ \addlinespace[2pt]
E13 & Which workflow said the information you needed more clearly? & Choice (Table~\ref{tab:categorical-comparisons}) \\ \addlinespace[2pt]
E14 & What one sentence did you most want to hear but did not hear? Name Workflow A or Workflow B. & Open response \\ \addlinespace[2pt]
E15 & In Workflow [A/B], did it feel like you were making the app together, not only watching from the side? & 1 to 5, co-building \\ \addlinespace[2pt]
E16 & When you tried the Workflow [A/B] app, what did you most want to figure out first? & Choice (Table~\ref{tab:categorical-conditions}) \\ \addlinespace[2pt]
E17 & In Workflow [A/B], did agent feedback change your confidence in your own judgment? & Choice (Table~\ref{tab:categorical-conditions}) \\ \addlinespace[2pt]
E18 & Does the finished app from Workflow [A/B] still feel like the problem you first wanted to solve? & Choice (Table~\ref{tab:categorical-conditions}) \\ \addlinespace[2pt]
E19 & During the process, did you feel the app drifted from your original idea? Was that drift more in Workflow A or Workflow B, and why? & Open response \\ \addlinespace[2pt]
E20 & After comparing Workflow A and Workflow B, which workflow had a special experience you did not have in the other workflow? & Choice (Table~\ref{tab:categorical-comparisons}) \\ \addlinespace[2pt]
E21 & If the previous answer was not \textquotedblleft{}no special difference\textquotedblright{}, what was that experience and how did you feel? Name Workflow A or Workflow B. & Open response \\ \addlinespace[2pt]
E22 & Overall, which workflow is closer to your ideal of a person and an agent making an app together? & Choice (Table~\ref{tab:categorical-comparisons}) \\ \addlinespace[2pt]
E23 & Anything else you want to add? (optional) Name Workflow A or Workflow B. & Open response \\ \addlinespace[2pt]
\bottomrule
\end{tabular}
\end{table*}

\subsection{Study Results}
\label{sec:study-results}

The 30 ideas spanned knowledge support, everyday services, productivity, creativity, finance, and health (Table~\ref{tab:app-themes}).

\begin{table}[htbp]
  \centering
  \caption{Themes of participant-built applications (30 distinct ideas).}
  \label{tab:app-themes}
  \small
  \begin{tabularx}{\linewidth}{>{\raggedright\arraybackslash}Xrr}
    \toprule
    Application theme & Count & Percentage \\
    \midrule
    Education, research, and knowledge support & 9 & 30.0\% \\
    Lifestyle, social interaction, and everyday services & 5 & 16.7\% \\
    Work and productivity & 5 & 16.7\% \\
    Entertainment and creativity & 4 & 13.3\% \\
    Finance and bookkeeping & 4 & 13.3\% \\
    Health and well-being & 3 & 10.0\% \\
    \bottomrule
  \end{tabularx}
\end{table}

\subsubsection{Core Comparisons}

Results are reported after converting each participant's A and B labels to the assigned conditions. For overall application preference, 20 participants chose AppLooper, 3 chose the baseline, and 7 rated them similarly. At the release decision, 16 reported greater trust in AppLooper's application, 3 in the baseline, and 11 expressed no clear preference. Both comparisons favored AppLooper after Holm correction ($p_H=.0020$ and $p_H=.0133$, Table~\ref{tab:categorical-owner-results}). Future workflow preference was significant only before correction ($p=.0266$, $p_H=.0532$), and ease of understanding was not significant ($p_H=.5034$). Responses expressing no clear preference remain in the reported totals but are excluded from the binomial tests. With all 30 responses retained, an exact trinomial test treating no clear preference as a third category gave $p_H=.0017$, $.0110$, $.0420$, and $.4352$ for preference, release trust, future workflow, and understanding. Dividing those responses equally between the conditions, with an odd response assigned to the baseline, gave $p_H=.0209$, $.1283$, $.1975$, and $.5847$. For future workflow preference, 16 chose AppLooper, 5 chose the baseline, and 9 would use either. For understanding, the corresponding counts were 12, 8, and 10 without a clear preference.

\begin{table*}[tbp]
  \centering
  \caption{Direct comparisons of participants' own applications (full sample $N=30$). No clear preference combines the responses about the same, either, and unclear. $p$ is the raw two-sided exact binomial test value for responses expressing a clear preference, and $p_H$ is Holm-adjusted across the four comparisons.}
  \label{tab:categorical-owner-results}
  \small
  \begin{tabularx}{\textwidth}{>{\raggedright\arraybackslash}Xrrrrr}
    \toprule
    Outcome & AppLooper & Baseline & \shortstack{No clear\\preference} & $p$ & $p_H$ \\
    \midrule
    Overall application preference & 20 & 3 & 7 & .0005 & .0020 \\
    Self-reported trust at the release decision & 16 & 3 & 11 & .0044 & .0133 \\
    Future workflow preference & 16 & 5 & 9 & .0266 & .0532 \\
    Ease of understanding & 12 & 8 & 10 & .5034 & .5034 \\
    \bottomrule
  \end{tabularx}
\end{table*}

Table~\ref{tab:all-paired-ratings} reports the 15 five-point rating items~\cite{likert1932technique,sullivan2013analyzing}. Mean final application quality was 3.800 for AppLooper and 3.167 for the baseline. The paired difference was .633, with a pointwise 95\% CI of [.200, 1.033] and $r_{\mathrm{rb}}=.561$ ($p=.0072$, $p_H=.1002$). Mean effort saved when using the finished application was 3.800 versus 2.867. The difference was .933, with a 95\% CI of [.400, 1.467] and $r_{\mathrm{rb}}=.618$ ($p=.0051$, $p_H=.0761$). Both were significant in unadjusted Wilcoxon tests, but neither survived Holm correction across 15 dimensions. Other mean directions favored AppLooper for core-task usability, overall effort saved, and problem resolution. Waiting and monitoring burden was slightly lower, and reading and selection burden slightly higher.

\begin{table*}[tbp]
  \centering
  \caption{Paired five-point rating outcomes (30 complete pairs per item). AppLooper and Baseline are condition means, and $\Delta$ is the AppLooper minus baseline mean difference. CIs are pointwise 95\% paired bootstrap intervals, and $r_{\rm rb}$ is the matched-pairs rank-biserial correlation. $p_W$ is the raw two-sided paired Wilcoxon value and $p_H$ applies Holm correction across all 15 ratings. Higher scores indicate greater burden for Q13 to Q16 and greater difficulty for Q21, and more positive evaluations for the other items. IDs are the item numbers in Tables~\ref{tab:study-instrument} and~\ref{tab:study-experience-instrument}.}
  \label{tab:all-paired-ratings}
  \small
  \setlength{\tabcolsep}{2.6pt}
  \begin{tabularx}{\textwidth}{@{}lXrrrrrr@{}}
  \toprule
  ID & Outcome & AppLooper & Baseline & $\Delta$ [95\% CI] & $r_{\rm rb}$ & $p_W$ & $p_H$ \\
  \midrule
  Q2 & Final application quality & 3.800 & 3.167 & \(0.633\) [0.200, 1.033] & \(0.561\) & 0.0072 & 0.1002 \\
  Q3 & Effort saved when using the finished application & 3.800 & 2.867 & \(0.933\) [0.400, 1.467] & \(0.618\) & 0.0051 & 0.0761 \\
  Q4 & Core-task usability & 3.733 & 3.200 & \(0.533\) [0.033, 1.033] & \(0.428\) & 0.0681 & 0.7487 \\
  Q10 & Overall effort saved & 3.733 & 3.167 & \(0.567\) [-0.033, 1.167] & \(0.348\) & 0.1174 & 1.0000 \\
  Q5 & Core-problem resolution & 3.633 & 3.300 & \(0.333\) [-0.167, 0.800] & \(0.251\) & 0.3090 & 1.0000 \\
  Q15 & Waiting and monitoring burden & 2.667 & 2.900 & \(-0.233\) [-0.667, 0.200] & \(-0.343\) & 0.2610 & 1.0000 \\
  Q13 & Reading and selection burden & 2.733 & 2.567 & \(0.167\) [-0.267, 0.633] & \(0.216\) & 0.4378 & 1.0000 \\
  Q14 & Context-recovery burden & 2.667 & 2.500 & \(0.167\) [-0.267, 0.600] & \(0.213\) & 0.4584 & 1.0000 \\
  Q16 & Verification and decision burden & 2.633 & 2.700 & \(-0.067\) [-0.467, 0.300] & \(-0.145\) & 0.7143 & 1.0000 \\
  Q20 & Perceived influence on changes & 3.667 & 3.400 & \(0.267\) [-0.033, 0.633] & \(0.515\) & 0.1304 & 1.0000 \\
  Q21 & Difficulty keeping up with changes & 2.900 & 2.733 & \(0.167\) [-0.100, 0.467] & \(0.359\) & 0.2655 & 1.0000 \\
  E1 & Overall satisfaction & 3.733 & 3.233 & \(0.500\) [0.033, 0.933] & \(0.489\) & 0.0550 & 0.6602 \\
  E4 & Traceability & 3.667 & 3.167 & \(0.500\) [0.167, 0.833] & \(0.714\) & 0.0169 & 0.2193 \\
  E7 & Ease of the workflow & 3.733 & 3.300 & \(0.433\) [-0.033, 0.867] & \(0.372\) & 0.1284 & 1.0000 \\
  E15 & Co-building & 3.367 & 3.400 & \(-0.033\) [-0.433, 0.367] & \(-0.033\) & 0.9301 & 1.0000 \\
  \bottomrule
  \end{tabularx}
\end{table*}

\subsubsection{Participant Development Feedback}

Twenty-six participants submitted 110 feedback messages across 41 workflows. AppLooper received 63 messages across 24 workflows and the baseline 47 across 17 workflows. Development conversations supplied 52 messages and trials 58. Before the recorded final owner judgment, precheck results appeared in 14 AppLooper workflows and owner-intent retest results in 20, with at least one of the two in 23.

Table~\ref{tab:owner-feedback} reports keyword categories for the 103 messages with recoverable text.

\begin{table*}[tbp]
  \centering
  \caption{Keyword categories of participant development feedback (recoverable messages $n=103$, 63 from AppLooper and 40 from the baseline).}
  \label{tab:owner-feedback}
  \small
  \begin{tabularx}{\textwidth}{>{\raggedright\arraybackslash}Xrrrr}
    \toprule
    Feedback category & AppLooper & Baseline & Total & Percentage \\
    \midrule
    Unmatched by keyword rules & 30 & 18 & 48 & 46.6\% \\
    Interface, visual design, and usability & 9 & 10 & 19 & 18.4\% \\
    Defects and interaction blockers & 11 & 5 & 16 & 15.5\% \\
    Content, data, and agent quality & 5 & 1 & 6 & 5.8\% \\
    General communication or nonspecific suggestions & 3 & 3 & 6 & 5.8\% \\
    Target users and personalization & 3 & 1 & 4 & 3.9\% \\
    Performance, waiting, and progress feedback & 2 & 2 & 4 & 3.9\% \\
    \bottomrule
  \end{tabularx}
\end{table*}

Process records illustrated links between feedback handling and release decisions. For a companion application, the owner requested free-text and preset companion roles and integration of interaction styles with companion types. Post-implementation retesting covered role options, free-text entry, and content persistence after leaving and reentering. All three checks passed, with corresponding screenshots retained. After this feedback item passed acceptance, the owner approved release during the subsequent inspection of the overall application.

For a research-news application, the owner asked, ``Why aren't there any sample articles in the daily frontier recommendations?'' The initial inspection found empty recommendations or loading failures, and the feedback item was returned for development. Post-revision retesting confirmed two articles explicitly labeled as synthetic examples, including titles and abstracts, and the feedback item passed acceptance. The owner nevertheless returned the overall application for revision at the subsequent release decision.

\subsubsection{Other Questionnaire Dimensions and Open-Ended Responses}

Across 30 experience reviews, mean traceability was 3.667 for AppLooper and 3.167 for the baseline ($p=.0169$, $p_H=.2193$). This item concerned clarity and confidence when recalling the application and version released or not released. For the same 30 participants, reading and selection burden was 2.733 versus 2.567 ($p=.4378$), context-recovery burden 2.667 versus 2.500 ($p=.4584$), and difficulty keeping up with changes 2.900 versus 2.733 ($p=.2655$). Co-building was 3.367 versus 3.400 ($p=.9301$). None of these four comparisons was significant before or after Holm correction ($p_H=1.0000$, Table~\ref{tab:all-paired-ratings}). None of the 15 scale comparisons survived Holm correction. These are exploratory findings.

Categorical responses are reported after each participant's A/B labels are mapped to the assigned conditions (Tables~\ref{tab:categorical-conditions} and~\ref{tab:categorical-comparisons}). Counts follow the option order in each row and include missing responses where applicable. Two participants received a questionnaire version that lacked the items on defects and on preference for the finished application. Participants described requirement drift in open responses.

\begin{table*}[tbp]
\small
\setlength{\tabcolsep}{3pt}
\renewcommand{\arraystretch}{1.13}
\caption{Condition-specific categorical responses (30 participants per condition). Counts follow option order. NA denotes an inapplicable option.}\label{tab:categorical-conditions}
\begin{tabular}{@{}p{.19\textwidth}p{.41\textwidth}p{.16\textwidth}p{.16\textwidth}@{}}
\toprule Item & Response options, in order & AppLooper & Baseline \\ \midrule
Q17 Most effortful stage & Reading and filtering / Context recovery / Waiting or active monitoring / Trial, acceptance, and decision / Owner-intent retesting / Other & 7 / 5 / 8 / 3 / 7 / 0 & 6 / 8 / 9 / 7 / NA / 0 \\ \addlinespace[3pt]
Q18 Perceived participation & Strong / Some / Weak / Almost none / Unclear & 9 / 15 / 5 / 1 / 0 & 7 / 11 / 9 / 2 / 1 \\ \addlinespace[3pt]
Q6 Reported stalls, errors, or blockers & None / One / Multiple / Unsure / Missing & 15 / 8 / 3 / 2 / 2 & 16 / 4 / 6 / 2 / 2 \\ \addlinespace[3pt]
E10 Hesitation at the release decision & Hesitated / Did not hesitate / Unclear & 8 / 19 / 3 & 10 / 19 / 1 \\ \addlinespace[3pt]
E11 Effect of agent messages & More confident / More hesitant / No difference / Did not notice / Other & 19 / 5 / 4 / 2 / 0 & 6 / 5 / 8 / 11 / 0 \\ \addlinespace[3pt]
E12 Most desired information & Current stage / Remaining time / Version changes / Problems encountered / When to try the application / Minimal information / Other & 6 / 7 / 7 / 3 / 6 / 1 / 0 & 7 / 7 / 4 / 4 / 7 / 1 / 0 \\ \addlinespace[3pt]
E16 Initial trial priority & Interface usability / Errors / Agent explanations / Requirement alignment / Other & 11 / 6 / 6 / 7 / 0 & 7 / 6 / 3 / 14 / 0 \\ \addlinespace[3pt]
E17 Effect of feedback on personal judgment & More confident judging independently / Less inclined to try it again personally / No effect / Unclear & 12 / 10 / 8 / 0 & 12 / 3 / 11 / 4 \\ \addlinespace[3pt]
E18 Resemblance to the original problem & Closely reflects the original problem / Somewhat different / Unclear & 12 / 15 / 3 & 14 / 11 / 5 \\ \addlinespace[3pt]
\bottomrule
\end{tabular}
\end{table*}

\begin{table*}[tbp]
\small
\setlength{\tabcolsep}{3pt}
\renewcommand{\arraystretch}{1.13}
\caption{Comparative categorical responses. Counts retain the original response categories, including uncertainty and neutral choices.}\label{tab:categorical-comparisons}
\begin{tabular}{@{}p{.28\textwidth}p{.49\textwidth}p{.15\textwidth}@{}}
\toprule Item & Response options, in order & Counts \\ \midrule
Q1 Overall application preference & AppLooper much better / AppLooper slightly better / About the same / Baseline slightly better / Baseline much better & 12 / 8 / 7 / 1 / 2 \\ \addlinespace[3pt]
Q11 Easier workflow to understand & AppLooper / Baseline / About the same / Unclear & 12 / 8 / 8 / 2 \\ \addlinespace[3pt]
Q22 Greater trust at release & AppLooper / Baseline / About the same / Unclear & 16 / 3 / 9 / 2 \\ \addlinespace[3pt]
Q24 Preferred workflow for another application & AppLooper / Baseline / Either / Neither & 16 / 5 / 9 / 0 \\ \addlinespace[3pt]
Q7 Preferred finished application for future use & AppLooper / Baseline / Either / Neither / Missing & 17 / 6 / 4 / 1 / 2 \\ \addlinespace[3pt]
E2 More efficient development workflow & AppLooper / Baseline / About the same / Unclear & 16 / 7 / 6 / 1 \\ \addlinespace[3pt]
E5 Clearer understanding of current activity & AppLooper / Baseline / About the same / Unclear & 15 / 5 / 8 / 2 \\ \addlinespace[3pt]
E8 More anxiety or impatience & AppLooper / Baseline / About the same / Unclear & 6 / 11 / 10 / 3 \\ \addlinespace[3pt]
E13 Clearer presentation of needed information & AppLooper / Baseline / About the same / Unclear & 14 / 6 / 9 / 1 \\ \addlinespace[3pt]
E20 Distinctive experience & AppLooper / Baseline / Both offered distinctive experiences / No particular difference & 12 / 6 / 5 / 7 \\ \addlinespace[3pt]
E22 Closer to ideal human--AI co-building & AppLooper / Baseline / About the same / Unclear & 16 / 7 / 6 / 1 \\ \addlinespace[3pt]
\bottomrule
\end{tabular}
\end{table*}

Asked whether the finished application still felt like the problem they first wanted to solve, 12 participants answered that the AppLooper application closely reflected it, 15 that it was somewhat different, and 3 were unclear, against 14, 11, and 5 for the baseline application. Open questions yielded 382 responses for analysis, 178 from application comparisons and 204 from overall-experience reviews.

\textbf{The agents' checks and visible changes.} A participant who rated both programming and AI tool experience as 0 said, ``I mostly observed the progress, so my sense of participation was weak,'' while valuing professional, effort-saving checks. A non-computing student who rated programming experience as 1 said, ``I made suggestions, and they had an effect,'' linking the revisions to a sense of control. Another valued AppLooper's checks but described the baseline application as ``more like something I planned entirely myself.''

\textbf{Requirement expansion.} The former inexperienced participant said AppLooper ``added some features that users needed but that I hadn't thought of initially.'' Another rating both experiences as 0 said it ``made some things that weren't requested,'' while acknowledging greater completeness. A computing student with programming experience 4 valued feature refinement and multiple perspectives, describing virtual-user backgrounds as ``much more varied than those of the users I had originally imagined.''

\textbf{Progress, waiting, and return.} One participant attributed efficiency to being able to ``clearly see how it made each change, why it made it, and what feedback there was.'' Another switched activities while waiting and returned to a page on which they could no longer proceed. A participant appreciated prompt issue responses but wished ``it could be a little faster'' and wanted to know ``how much changed in this version.''

None of the 12 associations between participant background and the outcomes was significant ($p\geq .4663$, adjusted $q\geq .9952$). The largest absolute correlation was between AI-tool experience and the quality difference ($\rho=-.138$).

\subsubsection{Workflow Progress and Owner Decisions}

Table~\ref{tab:workflow-outcomes} summarizes workflow records and owners' release judgments. The round-cap row counts each workflow once if it reached the 20-round cap (Section~\ref{sec:study-development-acceptance}) between its creation and the capture of its retained acceptance candidate. No retained acceptance candidate was recorded as meeting the readiness criteria. In AppLooper, 23 were frozen at the cap and 7 when errors stopped automatic progress. In the baseline, 25 were frozen after three rounds without new product changes and 5 at the cap. A round is one closed loop of revision and agent checks in AppLooper and one development iteration in the baseline, where the count restarts when a delivered candidate is reopened for revision. At the final owner judgment, AppLooper workflows had completed a median of 20 rounds (mean 17.5, range 1 to 20) and baseline workflows a median of 16.5 (mean 17.7, range 6 to 46). Elapsed times deduct recorded pauses and include queuing and periods when the owner was away.

\begin{table}[tbp]
  \centering
  \caption{Workflow outcomes (30 workflows per condition). Elapsed times are medians from workflow creation with recorded pauses deducted.}
  \label{tab:workflow-outcomes}
  \small
  \begin{tabularx}{\linewidth}{Xrr}
    \toprule
    Measure & AppLooper & Baseline \\
    \midrule
    First runnable version (minutes) & 10.1 & 11.0 \\
    Capture of retained acceptance candidate (hours) & 18.3 & 0.93 \\
    Recorded 20-round cap (workflows) & 23 & 5 \\
    Owner judged ready to release & 29 & 28 \\
    \quad of which recorded the 20-round cap & 23 & 4 \\
    Owner judged not yet ready to release & 1 & 2 \\
    \bottomrule
  \end{tabularx}
\end{table}

\subsubsection{Cumulative LLM Input and Output Volume}

We accumulated tokens, the units of text that a model reads and writes, from responses first appearing between workflow creation and the recorded final owner decision, deduplicating responses by session and message identifiers and excluding synthetic system messages. Cache-read input is input that the model service reuses from an earlier call, and cache-creation input is input stored for such reuse. Ordinary input, cache-read input, cache-creation input, and output were aggregated per workflow, and missing or all-zero records were not imputed. All 60 workflows had usage records, a creation time, and a final decision time, covering 1,680 session files. AppLooper and the baseline retained 57,250 and 6,640 responses with positive usage counts. Cache-creation input was zero throughout, and total input sums ordinary, cache-read, and cache-creation input. Table~\ref{tab:development-duration} reports per-workflow means and medians.

\begin{table*}[htbp]
  \centering
  \caption{Cumulative LLM input and output volume from invocation records, with 30 workflows per condition. $M$ and $Md$ denote the per-workflow mean and median. Responses are counted individually, and cumulative tokens are reported in millions. Total input includes ordinary and cached input.}
  \label{tab:development-duration}
  \small
  \begin{tabularx}{\textwidth}{>{\raggedright\arraybackslash}Xrrrr}
    \toprule
    & \multicolumn{2}{c}{AppLooper} & \multicolumn{2}{c}{Control baseline} \\
    \cmidrule(lr){2-3}\cmidrule(lr){4-5}
    Metric & $M$ & $Md$ & $M$ & $Md$ \\
    \midrule
    Model responses with positive usage counts & 1908.333 & 1854.500 & 221.333 & 157.000 \\
    Ordinary input & 36.242 & 33.915 & 2.995 & 1.993 \\
    Cache-read input & 130.363 & 107.433 & 12.219 & 6.256 \\
    Output & 1.150 & 1.210 & 0.132 & 0.096 \\
    Total input & 166.605 & 140.023 & 15.213 & 8.225 \\
    \bottomrule
  \end{tabularx}
\end{table*}

Mean ordinary input was 36.242 versus 2.995 million tokens, output 1.150 versus 0.132 million, and total input including caches 166.605 versus 15.213 million, for AppLooper and the baseline respectively. These totals sum retained usage records within the aggregation window.

\section{Discussion}

This section interprets the study results with respect to owner preference, release trust, and participation during development, then describes the limitations of the study and directions for future work.

\subsection{Interpreting the Study Results}

\textbf{Overall preference and release decisions.} The two significant comparisons concerned overall application preference and self-reported trust at release (Table~\ref{tab:categorical-owner-results}). The reasons participants gave for their trust point to two things they could observe. One was that parties other than the developing agent had checked the version. A participant with no programming experience felt ``more at ease, since other agents have helped verify it,'' and another found that ``development and revision over multiple rounds feels more trustworthy.'' The other was seeing their own words take effect. A participant who rated programming experience as 1 reported more confidence in the finished product because ``when I told it how to change something, it did change in the direction I said.'' Neither reason involves reading code. Owners thus appear to form release trust from who has checked a version and from whether their own requests visibly took effect, the two kinds of evidence that AppLooper records with each candidate version. In line with this reading, 19 participants reported that the agents' messages made them more confident in AppLooper, against 6 in the baseline (Table~\ref{tab:categorical-conditions}), and 15 of those 19 also named AppLooper as the workflow whose application they trusted more at release.

\textbf{Product experience and development burden.} Paired mean differences in final application quality, effort saved when using the finished application, and traceability were .633, .933, and .500. These positive estimates provide exploratory evidence. Effort saved in using the finished product and effort saved during construction concern distinct activities, and the latter comparison had an unadjusted $p=.1174$ (Table~\ref{tab:all-paired-ratings}). Participants' accounts locate the construction effort they were spared in repeated testing. One wrote of AppLooper that ``after one instruction, several agents verify it,'' which spared the check that otherwise followed every instruction, and another described the baseline as requiring them, besides stating needs, to keep testing the application and proposing changes. After the agents carry out the checks, the owner's remaining work is to judge the issues they report and the revisions that follow. The loop therefore changes the kind of work an owner does during construction, from finding defects to deciding on reported ones.

\textbf{Staying informed and resuming participation.} Fifteen participants named AppLooper as the workflow that made it clearer what was happening and 5 named the baseline, and 14 against 6 judged that it stated the information they needed more clearly (Table~\ref{tab:categorical-comparisons}). Visible activity itself carried information. One participant wrote of the baseline that ``because I could not see status updates, I wondered whether it had crashed,'' and of AppLooper that watching the agents interact showed where an unclear description of theirs needed refining. Asked for the one sentence they most wanted to hear, participants gave answers of three kinds. Some asked where the work stood, such as which step each workflow had reached or how much time was still needed. Some asked what had changed and why, such as ``how much changed in this version'' or a statement that a given change had been made according to a given request of theirs. Others asked what they should do next, as in ``do I need to run the program through myself?'' The choice item on the most desired information was similarly spread across the current stage, remaining time, version changes, and when to try the application, with no option chosen by more than 7 participants for either workflow (Table~\ref{tab:categorical-conditions}). Taken together, these responses indicate that an owner returning after an absence needs to know about progress, about changes and their causes, and about the next action expected of them.

\textbf{Delegation and agency.} The two conditions had similar mean co-building ratings, and 24 participants reported strong or some participation in AppLooper against 18 in the baseline (Table~\ref{tab:categorical-conditions}). Their explanations show two measures of participation. Some participants counted the work they did by hand. One with no programming experience felt less involved in AppLooper because the agents completed testing and optimization once the request was stated, and more involved in the baseline ``because I had to do the trials.'' Others counted the effect of what they said. A participant who rated programming experience as 3 felt very involved in AppLooper because several roles with a clear division of work examined many details, and another contrasted suggestions that had an effect in AppLooper with suggestions in the baseline that were not acted on. Checks carried out by the agents lower participation on the first measure and can raise it on the second, which is consistent with similar mean ratings alongside differing individual accounts. From the two measures above, once the agents carry out the checks, the owner's agency comes from comments the owner makes and can see taking effect, and showing which revisions and checks followed a comment makes that contribution recognizable. Preferences varied even among similarly experienced participants, and none of the 12 background associations was significant. One participant remarked that a quick result is wanted on some occasions and explicit attention to details on others. Owners therefore need flexibility to observe, comment, or test according to their task and preferences.

\textbf{Negotiating requirements and acceptance.} Among two participants with no programming or AI tool experience, one welcomed additional features, while another noted unrequested additions alongside a more complete implementation. A single participant could hold both views. One who rated programming experience as 0 favored the AppLooper application for being ``able to add parts I had not asked for'' and also reported that it had made extra buttons that were not requested. Additions were welcomed when they met a need the participant recognized and were questioned when they appeared as interface elements the participant had not asked for. Of the 15 participants who judged the AppLooper application somewhat different from the problem they first wanted to solve, 8 preferred it, 4 rated the two applications about the same, and 3 preferred the baseline application. Departure from the initial idea was therefore compatible with preferring the result, and the owner's response to an addition depends on whether the owner can recognize the need it serves. The process cases also distinguish the retest of a single feedback item from acceptance of the whole application. The companionship application received release approval after feedback retesting, whereas the research-news application was returned despite passing its retest. A passed retest of a single feedback item and release approval were therefore not necessarily linked. This suggests that the owner's release judgment may have drawn on considerations beyond that retest.

\subsection{Limitations}

\textbf{Study scope and participation.} The 30 participants primarily evaluated applications generated from their own ideas through the initial release decision. Participants themselves chose when to try applications and how much feedback to provide, with human feedback reaching 41 of the 60 workflows. The category counts in Table~\ref{tab:owner-feedback} therefore reflect both the issues owners perceived in an application and their willingness to report them. Limited server resources caused network and memory problems that affected the study to some extent. Research safeguards constrained external services and real-data access. Some applications that owners conceived as online services ran on synthetic data rather than on data accumulated through use by their intended users. The sample and variation in participation constrain comparisons across participant backgrounds and application types, as well as generalization to sustained use by actual target users.

\textbf{Study design and attribution.} The study measured application owners' evaluations of the applications they constructed. Those evaluations therefore reflect participation during development and the experience of the application awaiting release, among other factors. Owners rated application quality, and visible condition differences could influence expectations and responses~\citep{dell2012yours}. Evaluating trust calibration, the match between an owner's trust and the reliability of the application, requires evidence about perceived trust and appropriate reliance~\citep{lee2004trust,grunde2026overseeing}. The study did not collect assessments of the final applications' experience and quality from the users the owners intended to serve. Such assessments would be independent of the owner's participation in development and therefore more objective.

\textbf{Control condition.} Claude Code Loop is an ablation on the same iterative coding base. It compares the key mechanisms AppLooper adds, rather than another heterogeneous best implementation for this task. On the other hand, the framework implemented in this study is not a theoretically optimal framework. Its modules have potentially better alternatives, which further indicates that its effects still have room to improve.

\textbf{Trade-offs reported by owners.} The AppLooper advantages were concentrated in overall preference and release trust. Among these 30 participants, the means for reading and selection burden (2.733 versus 2.567, $p=.4378$), context-recovery burden (2.667 versus 2.500, $p=.4584$), and difficulty keeping up with changes (2.900 versus 2.733, $p=.2655$) were higher for AppLooper, and co-building was 3.367 versus 3.400 ($p=.9301$). None of these four comparisons was significant before or after Holm correction ($p_H=1.0000$). Difficulty keeping up with changes may have several sources. AppLooper increased the number of revisions and role-specific checks available for inspection, each recorded with its version, while the complexity of multiple agents can make concurrent changes harder to follow. Ease of understanding showed no significant difference. None of the 15 rating comparisons survived Holm correction, and these ratings do not show that resumption after an absence was eased. Some participants felt less involved when checks were delegated to agents, some noted additions they had not requested, and some described AppLooper as slower to respond than the baseline. These accounts indicate that checks by the agents can reduce the owner's verification work while also reducing the owner's sense of authorship and increasing waiting.

\textbf{Development time, computation, and resource measurement.} The retained AppLooper acceptance candidate was captured a median of 18.3 hours after workflow creation, against 0.93 hours for the baseline (Table~\ref{tab:workflow-outcomes}), and AppLooper workflows used about 11 times the total input tokens (166.6 versus 15.2 million per workflow, Table~\ref{tab:development-duration}). The applications that participants preferred and trusted more were therefore produced with more elapsed time and computation, which limits attribution of the observed differences to workflow design. Waiting and monitoring burden was 2.667 versus 2.900 ($p=.2610$), 6 participants named AppLooper as the workflow that made them more anxious or impatient and 11 named the baseline, and 16 judged AppLooper the more efficient development workflow and 7 the baseline (Table~\ref{tab:categorical-comparisons}). A longer run therefore did not by itself make the waiting feel heavier, and that feeling may have depended more on whether owners still had to stay and check it themselves.

\subsection{Future Work}

\textbf{Intended-user feedback in continued release.} After the actual users an application is intended to serve have been fully informed and have consented, their feedback across successive released versions can become part of the loop and enter continued optimization. That optimization can also add personalized learnable components, so that feedback from actual users takes part in optimizing each component and the system comes closer to those users' preferences and needs. Issue reports and experience judgments from use are recorded with the version they concern and proceed into the subsequent revision, retesting, and next release. Task completion, reported experience, and whether later versions address the reported feedback can assess this process across versions.

\textbf{Adaptive orchestration and resource efficiency.} The present loop fixes some orchestration choices in advance and leaves others to the model that receives the task context. The deterministic scope policy, feedback routing, the fixed test sequence after each revision, owner-intent retesting of at most two rounds, and the pause for acceptance under preset conditions belong to the first group. The second-stage admit, rewrite, or reject decision, persona and critical-journey generation, defect interpretation, and abstention judgments are produced by the model from the supplied context. The resulting inspection depth, retesting frequency, and role order have not been optimized as a weighted tradeoff among expected effect, sparse and low-burden owner intervention, and token consumption. The tradeoff among the cost of owner intervention, effect, and efficiency can continue to be optimized through continued interaction, toward a balance that adjusts as the interaction proceeds. Continued interaction could collect effect signals for whether a check changed requirement coverage, defect status, or later acceptance, intervention signals for when the owner was asked to act and for the reading and trial burden of that request, and token signals recorded by role and retest on the candidate version that incurred them. Capturing these signals and using them jointly remains an open problem. Reinforcement learning, which improves a decision policy from the observed outcomes of its earlier choices, can update this orchestration from the combined signal during continued interaction~\citep{kaelbling1996reinforcement} and extend the applicability of the associated algorithms and the framework.

\textbf{Post-release evolution and virtual-user calibration.} After an application is released, and following ethical review and informed consent, de-identified usage records can be used to adjust the virtual users. In the to-do application example, if actual users report from a new form of terminal that a deadline cannot be chosen, that case can be added to the virtual users' tasks and usage habits, and users of that terminal can be added to the cohort. Virtual users can then screen reproducible common problems first. Owners can focus on newly stated requirements and on problems for which a virtual user's check result is inconsistent with actual users' feedback. For example, the check can indicate that a deadline can be chosen on that terminal, and actual users can still report that it cannot. Where repeating the same action still leaves it undetermined whether an experience matches actual users' expectations, for example whether the reminder interval shown on a to-do application's card leaves actual users enough time to prepare, the owner can also make that judgment.

\section{Conclusion}

When coding agents revise an application across successive rounds, checking of each version is often limited to the agent's own reports and the owner's manual inspection. This constrains the problems found before release, and owners still have limited support for working out what has changed in their absence, assessing the current application, and judging whether a reported completion is accurate. We proposed AppLooper, a multi-role agentic application development loop for owner participation and feedback from hands-on use. In AppLooper, owners try the application and give feedback, agents in several roles check, revise, and retest it, and owners make the final acceptance decision. The evidence from these steps is recorded with the version it concerns, giving owners grounds for trusting the application they release. In a paired study with 30 participants, overall application preference and self-reported trust at the release decision favored AppLooper after Holm correction ($p_H=.0020$ and $.0133$). Participants described valuing the agents' checks and visible changes, while differing in how closely they wanted to shape new features and follow development.

\section*{Use of Generative AI}

The authors used ChatGPT (OpenAI) to draft selected sentences from their ideas and descriptions, assist with English translation, edit and check the manuscript, support literature searching and screening, edit some modules in some of the figures, and review interpretations of the study findings. ChatGPT image generation (OpenAI) produced the role illustrations in Figures~\ref{fig:applooper-leading} and~\ref{fig:framework} from author descriptions. ChatGPT (OpenAI) assisted with building the system code. The authors reviewed and revised the outputs and take full responsibility for the integrity and accuracy of the study data, cited sources, and conclusions.

\bibliographystyle{ACM-Reference-Format}
\bibliography{sample-base}
\end{document}